\documentclass[12pt,draftclsnofoot,12pt,onecolumn]{IEEEtran}
\usepackage[latin9]{inputenc}
\usepackage{amsthm}
\usepackage{amsmath}
\usepackage{amssymb}
\usepackage{graphicx}
\usepackage{epstopdf}

\makeatletter

\providecommand{\tabularnewline}{\\}

\theoremstyle{plain}
\newtheorem{thm}{\protect\theoremname}
\theoremstyle{plain}
\newtheorem{prop}[thm]{\protect\propositionname}
\theoremstyle{plain}
\newtheorem{cor}[thm]{\protect\corollaryname}
\theoremstyle{remark}
\newtheorem{rem}[thm]{\protect\remarkname}

\usepackage[T1]{fontenc}
\providecommand{\corollaryname}{Corollary}
\providecommand{\theoremname}{Theorem}

\providecommand{\corollaryname}{Corollary}
\providecommand{\theoremname}{Theorem}

\providecommand{\corollaryname}{Corollary}

\providecommand{\theoremname}{Theorem}

\providecommand{\corollaryname}{Corollary}

\providecommand{\theoremname}{Theorem}

\providecommand{\corollaryname}{Corollary}
\providecommand{\remarkname}{Remark}
\providecommand{\theoremname}{Theorem}

\providecommand{\corollaryname}{Corollary}
\providecommand{\remarkname}{Remark}
\providecommand{\theoremname}{Theorem}

\providecommand{\corollaryname}{Corollary}
\providecommand{\remarkname}{Remark}
\providecommand{\theoremname}{Theorem}

\providecommand{\corollaryname}{Corollary}
\providecommand{\propositionname}{Proposition}
\providecommand{\remarkname}{Remark}
\providecommand{\theoremname}{Theorem}

\makeatother

\providecommand{\corollaryname}{Corollary}
\providecommand{\propositionname}{Proposition}
\providecommand{\remarkname}{Remark}
\providecommand{\theoremname}{Theorem}

\begin{document}

\title{Optimal Primary-Secondary user Cooperation Policies in Cognitive
Radio Networks}

\author{Nestor~D.~Chatzidiamantis,~\IEEEmembership{Member,~IEEE},~Evangelia~
Matskani, Leonidas~Georgiadis, \IEEEmembership{Member,~IEEE}, Iordanis
Koutsopoulos,~\IEEEmembership{Member,~IEEE}, and~Leandros Tassiulas,~\IEEEmembership{Fellow,~IEEE}%
\thanks{Part of this paper has been presented in WiOpt 2014.%
}%
\thanks{This research has been co-financed by the European Union (European
Social Fund \textendash{} ESF) and Greek national funds through the
Operational Program ``Education and Lifelong Learning\textquotedbl{}
of the National Strategic Reference Framework (NSRF) - Research Funding
Program: Thales. Investing in knowledge society through the European
Social Fund.%
}%
\thanks{N. D. Chatzidiamantis, E. Matskani and L. Georgiadis are with the
Department of Electrical and Computer Engineering, Aristotle University
of Thessaloniki, Greece, (e-mails:\{nestoras,matskani,leonid\}@auth.gr).%
}%
\thanks{I. Koutsopoulos is with Department of Informatics, Athens University
of Economics and Bussiness (AUEB), Greece, and the Centre for Research
and Technology Hellas (CERTH), Greece, (e-mail: jordan@aueb.gr).%
}%
\thanks{L. Tassiulas is with the Department of Computer and Communications
Engineering, University of Thessaly, Greece, and the Centre for Research
and Technology Hellas (CERTH), Greece, (e-mail: leandros@uth.gr).%
} }
\maketitle
\begin{abstract}
In cognitive radio networks, secondary users (SUs) may cooperate with
the primary user (PU) so that the success probability of PU transmissions
are improved, while SUs obtain more transmission opportunities. However,
SUs have limited power resources and, therefore, they have to take
intelligent decisions on whether to cooperate or not and at which
power level, in order to maximize their throughput. Cooperation policies
in this framework require the solution of a constrained Markov decision
problem with infinite state space. In our work, we restrict attention
to the class of stationary policies that take randomized decisions
of an SU activation and its transmit power in every time slot based
only on spectrum sensing. Assuming infinitely backlogged SUs queues,
the proposed class of policies is shown to achieve the maximum throughput
for the SUs, while significantly enlarging the stability region of
PU queue. The structure of the optimal policies remains the same even
if the assumption of infinitely backlogged SU queues is relaxed. Furthermore,
the model is extended for the case of imperfect channel sensing. Finally,
a lightweight distributed protocol for the implementation of the proposed
policies is presented, which is applicable to realistic scenarios.\end{abstract}


\section{Introduction}

Cognitive radio networks (CRNs) have received considerable attention
due to their potential for improving spectral efficiency \cite{FCC,Mitola,A:Haykin}.
The main idea behind CRNs is to allow unlicensed users, also known
as\textit{ secondary users} (SU), to identify temporal and/or spatial
spectrum ``holes'', i.e., vacant portions of licensed spectrum,
and transmit opportunistically, thus gaining access to the underutilized
shared spectrum while maintaining limited interference to the licensed
user, also known as \textit{primary user} (PU). This communication
paradigm has been referred to as ``Dynamic Spectrum Access'' (DSA)
in the technical literature \cite{A:DSA1,DSA2}.

Much prior work on DSA CRNs has been focused on the problem of optimal
spectrum assignment to multiple SUs \cite{Peng,Chen-2008-ID52,Urgaonkar-2009-ID47}.
Several resource allocation algorithms have been proposed, based on
either the knowledge of PU transmissions obtained from perfect spectrum
sensing mechanisms \cite{Peng} or from a probabilistic maximum collision
constraint with the PUs \cite{Chen-2008-ID52}. Of particular interest
is the opportunistic scheduling policy for SUs suggested in \cite{Urgaonkar-2009-ID47},
which maximizes SUs' throughput utility while guarantees low number
of collisions with the PU, as well. In all these works it is assumed
that no interaction between PUs and SUs exists.

Recently, the concept of cooperation between PU and SUs in CRNs emerged,
as a means for providing benefits for both types of users. These benefits
stem from the fact that, by exploiting the transmit power resources
of SUs towards improving the effective transmission rate of the PU,
the chances that the PU queue will be empty are increased, and hence
the PU channel is free to use more often.

From an information theoretic perspective, cooperation between SUs
and PUs at the physical layer has been investigated in many works
(see \cite{Goldsmith-2009-ID359} and references therein). Queuing
theoretic aspects and spectrum leasing strategies for cooperative
CRNs have been investigated in \cite{Simeone-2008-ID472,Simeone-2007-ID29,Krikidis-2009-ID233,Kompella2011,Neely_cooper}.
Specifically, spectrum leasing strategies where the PU leases a portion
of its spectrum to SUs in return for cooperative relaying were suggested
in \cite{Simeone-2008-ID472}. A protocol where a SU relays the PU
packets that have not been correctly received by their destination,
was suggested and investigated in terms of SU stable throughput in
\cite{Simeone-2007-ID29}, while similar protocols were suggested
and compared in \cite{Krikidis-2009-ID233}, considering various physical
layer relaying strategies. In \cite{Kompella2011}, the performance
of a specific class of PU-SU cooperation policies was investigated
in terms of PU and SU stable throughput, assuming that SU is allowed
to transmit simultaneously with the PU, even if the PU is busy. 

In this work we study optimal cooperative PU-SUs transmission control
algorithms with the objective to make as efficient use of the PU channel
as possible, namely maximize a function of the transmission rates
of the SUs, while guaranteeing unobstructed packet transmission for
the PU, and stability of its queue. SUs have limited transmit power
resources, therefore intelligent cooperation decisions must be taken.
This is the main idea behind the work in \cite{Neely_cooper}, where
a dynamic decision policy for the SUs activities (i.e., whether to
relay PU transmissions and at which power level) is suggested. The
proposed policy is proved to be optimal, however, its basic requirement
is that the PU packet arrival rates must be lower than a threshold
value, which guarantees that the PU queue is stable even when SUs
never cooperate. This regime places significant restrictions on the
achievable PU stability region, since the sustainable arrival rates
of PUs may be much larger than this threshold value.

We present policies that significantly increase the range of PU arrival
rates for which PU-SUs cooperation can be beneficial. Specifically,
we investigate transmission policies for cooperative CRNs that can
be applied even when PU transmission rates are above the threshold
set by \cite{Neely_cooper}, while still permitting the SUs to utilize
the channel for their own transmissions. Since the SU decision options
and success probabilities are different during the idle and busy PU
periods, while the PU queue size is in turn affected by the cooperation
decisions, such policies require in general the solution of a non-trivial
constrained Markov decision problem with infinite state space, where
the state is the size of the PU queue. The solutions for such Markov
decision problems suffer from large convergence times and their implementation
in general requires knowledge of the PU queue size \cite{B:Altman}.

The main contributions of this work are summarized as follows. 
\begin{enumerate}
\item We introduce a class of stationary policies which take random decisions
on SU activities in every time-slot based only on the PU channel spectrum
sensing result, i.e., the PU channel being busy or idle. The proposed
class of policies is applicable when either SUs are infinitely backlogged
or a general SU packet arrival process is assumed. The benefits of
our approach are as follows. First, our approach is proven to achieve
the same set of SU rates as the more general policies in which ($i$)
decision may depend on the PU queue size, or ($ii$) a SU packet may
be transmitted instead of a PU packet when the PU queue is non-empty.
Hence, the policies in the proposed class of stationary ones are sufficient
for optimality with respect to any utility function. Second, compared
to other policies, it allows for a significantly larger range of PU
traffic arrival rates for which the PU queue is stable, thus increasing
the PU throughput. Even more interestingly, the enlargement of the
PU stability region still allows the SUs to utilize slots that are
unused by the PU, in order to transmit their own traffic. Finally,
as long as the system parameters remain the same, the decision variables
associated with our policy may be computed offline, through solving
a convex optimization problem via efficient interior point methods,
and can be used to realize the policy in real-time. 
\item Since the proposed policies are based solely on the PU channel state
sensing result, we also investigate the effects of imperfect spectrum
sensing mechanism in their performance. Considering this case, we
incorporate all possible sources of errors and inefficiencies in our
model and describe the new performance space of the proposed policies.
However, when channel sensing errors are introduced, the determination
of the associated control variables requires the solution of a non-convex
optimization problem and the optimal solution becomes hard to determine. 
\item A distributed implementation of the proposed cooperation policies,
applicable to the case of concave SU utility functions, is designed,
which is based on a decentralized computation of the problem control
variables via the alternating direction method of multipliers. This
version offers a robust alternative to the centralized implementation
and distributes the computational burden across network nodes without
loss in performance. 
\end{enumerate}
The remainder of the paper is organized as follows. In Section \ref{sec:System-Model},
we introduce the system model. In Section \ref{sec:Prformance-regions-identical}
we describe the mode of operation of the proposed restricted class
of randomized policies and show their optimality. Exogenous packet
arrivals to SU queues and the effects of imperfect spectrum sensing
mechanism are investigated in section \ref{sec:Extensions}. The distributed
implementation of the proposed class of policies is developed in Section
\ref{sec:Distributed-Implementation}. Section \ref{sec:Simulation-Results}
presents simulation results and finally, concluding remarks are provided
in Section \ref{sec:Conclusions}.

\section{\label{sec:System-Model}System Model}

\begin{figure}
\centering\includegraphics[width=0.4\columnwidth]{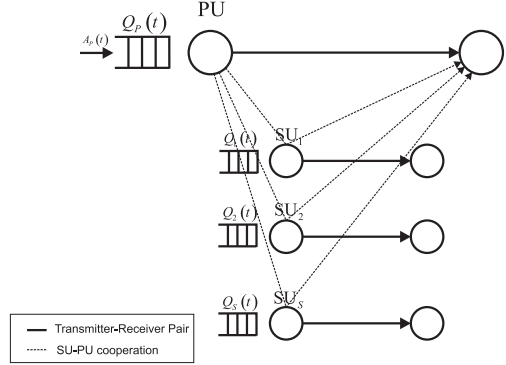}

\protect\caption{The system model under consideration.}
\label{Fig:system_model}
\end{figure}

We consider the system model with one PU and multiple SUs depicted
by Fig. \ref{Fig:system_model}. Specifically, the PU is the licensed
owner of the channel and transmits whenever it has data to send. On
the other hand, SUs do not have any licensed spectrum and seek transmission
opportunities on the PU channel. We assume that one%
\footnote{The presented analysis can be applied in cases where more than one
SUs can cooperate with PU, by replacing the selected SU by a subset
of SUs. %
} of the SUs can cooperate with the PU in order to improve the success
probability of PU transmissions. This can be achieved by allocating
part of the SU power resources towards that purpose. In practice,
SU cooperation may be realized with various techniques that span one
or more communication layers. For example, the SU may relay PU traffic
(e.g. through decode-and-forward, or amplify-and-forward) \cite{Neely_cooper}.
Alternatively, this aid by the SU can be provided by means of link
layer techniques, such as retransmission of the overheard PU packet
by the SU, or even through physical layer techniques (e.g. simultaneous
transmission of the PU packet by the SU, in order to improve the signal-to-interference-plus-noise
ratio at the PU receiver) \cite{Krikidis-2009-ID233}. The model is
transparent to capture the generality of all these techniques, all
of which are factored in the problem in terms of the SU consumed transmit
power resources.

Furthermore, after sensing the PU channel, SUs decide on which SU
will cooperate so as to transmit PU data and at which power level
(if the PU channel is busy), or which SU will transmit its own data
and at which power level (if the PU channel is idle). In what follows
we describe the parameters of the system model under consideration
as well as the available controls.

\subsection{\label{sub:System-Model-Parameters}System Model Parameters}

We consider the time-slotted model, where time slot $t=0,1,...$ corresponds
to time interval $[t,t+1)$; $t$ and $t+1$ are called the ``beginning''
and ``end'' of slot $t$ respectively. The PU queue receives new
packets in each time slot $t$ according to an independent and identically
distributed (i.i.d.) arrival process $A_{p}\left(t\right)$ with mean
rate $\mathbb{E}\left[A_{p}(t)\right]=\lambda_{p}$ packets/slot and
$\mathbb{E}\left[\left(A_{p}\left(t\right)\right)^{2}\right]<\infty$.
We assume that the SUs are backlogged so that they \emph{always} have
packets to transmit.

We denote by $\mathcal{S}$ the set of SUs. Each SU $s\in{\cal S}$
can transmit using one of $I_{s}$ power levels, $P_{s}\left(i\right)$,
$i=1,...,I_{s}$, where $P_{s}(i)<P_{s}(i+1)$. To simplify the description
that follows, we set $P_{s}(0)=0$. An SU $s$ may use any of these
power levels to either transmit its own data or to assist the PU as
discussed above. At each time slot, only a single packet transmission
can take place. Furthermore, when transmission of packets from the
PU takes place, at most one of the SUs can cooperate. There is a constraint
on the long-term average power $\hat{P}_{s}$ consumed by each $s\in{\cal S}$.
Hence, for every $s\in{\cal S}$, if $i(t)$ is the power level used
by $s$ at slot $t$, it must hold, 
\begin{equation}
\limsup_{t\rightarrow\infty}\frac{1}{t}\sum_{\tau=0}^{t}\mathbb{E}\left[P_{s}\left(i\left(\tau\right)\right)\right]\leq\hat{P}_{s},\,\, i\left(\tau\right)\in{\cal I}_{s}^{0},\label{eq:power_constraint}
\end{equation}
where $\mathbb{E}\mbox{\ensuremath{\left[\cdot\right]}}$ denotes
expectation, ${\cal I}_{s}=\left\{ 1,2,...,I_{s}\right\} $ and ${\cal I}_{s}^{0}={\cal I}_{s}\cup\left\{ 0\right\} $.

We assume an erasure channel model, i.e., that each transmission (by
the PU or one of the SUs) is either received correctly or erased. 
\begin{itemize}
\item When SU $s$ transmits one of its own packets with $i$th power level,
$i\in\mathcal{I}_{s}^{0}$, the probability of success is $r_{s}(i)$,
where $r_{s}(0)=0,$ i.e. the success probability is zero if no power
is used for transmission. 
\item When SU $s$ cooperates with the PU, (namely it assists in the transmission
of PU packets by transmitting with $i$th power level), the success
probability of the PU transmitted packet is $r_{p}\left(s,i\right).$
If $i=0$, the SU ``cooperates'' with zero transmission power, hence
in effect no cooperation takes place; therefore it is natural to assume
that $r_{p}(s,0)=r_{p}(0)\geq0$ for all $s\in\mathcal{S}$, where
$r_{p}(0)$ denotes the probability of successful packet transmission
by the PU when the SUs do not cooperate. In addition, we assume that
$r_{p}\left(s,i\right)\leq r_{p}\left(s,i+1\right)$, i.e., the probability
of successful reception is a non-decreasing function of transmission
power. 
\end{itemize}

\subsection{\label{sub:Control-Decisions}Available Controls }

In the beginning of time slot $t$ there are various control options,
depending of the status of the primary queue $Q_{p}\left(t\right)$.
In case $Q_{p}\left(t\right)>0$ (namely, the PU channel is busy),
then the available controls are: 
\begin{itemize}
\item A packet from the PU queue is transmitted, and transmission of SU
packets is excluded. We refer to this constraint as the \emph{PU priority
constraint}\textit{.} 
\item A SU $s$ is selected for cooperation with the PU in order to assist
the transmission of the PU packet. 
\item The $i$th power level, $i\in\mathcal{I}_{s}^{0}$, is selected, so
that $s$ cooperates with the PU using power level $P_{s}(i)$. When
$i\left(t\right)=0$ no cooperation takes place. 
\end{itemize}
On the other hand, when $Q_{p}\left(t\right)=0$ (namely, the PU channel
is idle), the available controls are the following: 
\begin{itemize}
\item A SU $s$ is selected to transmit its own packet. 
\item The $i$th power level, $i\in\mathcal{I}_{s}^{0}$, is selected, so
that $s$ transmits its own packets using power level $P_{s}(i)$.
If $i=0$, no transmission takes place in slot $t$. 
\end{itemize}

\subsection{\label{sub:Control-Objective}Admissible Policies, Rate Region, Performance
Objective and Extended class of Policies}

A control policy is called \emph{admissible} if the following policy
constraints are satisfied: 
\begin{itemize}
\item PU priority constraint is satisfied. 
\item The PU queue must be mean-rate stable, i.e., the output long-term
average rate of the PU queue should be equal to its long term average
input rate \cite{B:Neely_book}. 
\item The average power constraints of (\ref{eq:power_constraint}) are
satisfied. 
\end{itemize}
Under an admissible policy, each SU $s\in{\cal S}$ obtains a long-term
average transmission rate equal to 
\begin{equation}
\bar{r}_{s}=\lim_{t\rightarrow\infty}\inf\frac{\sum_{\tau=0}^{t-1}\mathbb{E}\left[r_{s}(P_{s}(i(\tau))\right]}{t}\label{eq:aver_rate}
\end{equation}
where $P_{s}\left(i\left(t\right)\right)$ is the power level at which
$s$ transmits in slot $t$. In the sequel, we denote by $\mathbf{\bar{r}}$
the vector of the long-term average transmission rates of SUs, i.e.,
$\mathbf{\bar{r}\triangleq}\left\{ \bar{r}_{s}\right\} _{s\in\mathcal{S}}$.
The \emph{achievable rate region} for the problem under consideration
is defined as the set of vectors of SU rates $\mathbf{\bar{r}}$ that
can be obtained by all admissible policies.

The selection of an admissible policy depends on the particular optimization
objective, which is expressed as a function of the vector of achievable
long-term average SU transmission rates $\mathbf{\bar{r}}$. The optimization
objective is of the form: 
\begin{equation}
\mbox{maximize: }f\left(\mathbf{\bar{r}}\right)\label{eq:function_aver_rate}
\end{equation}
where $\mathbf{\bar{r}}$ belongs to the rate region. In the simplest
case, $f\left(\cdot\right)$ is a linear function of $\mathbf{\bar{r}}$,
however, fairness considerations may require $f\left(\cdot\right)$
to be a nonlinear (usually separable) function of $\mathbf{\bar{r}}$,
\cite{Srikant2013}, \cite{A:Mo}.

The PU queue size $Q_{p}(t)$ can be seen as the state of a constrained
Markov Decision Process problem \cite{B:Altman}, where the constraints
are imposed by the policy constraints described above. Let $\mathcal{C}_{1}$
be the class of admissible policies of this Markov Decision Process.
This class contains policies that are based on past history actions
and includes the class of randomized stationary policies of the following
form: 
\begin{itemize}
\item When $Q_{p}(t)=m,\, m>0$, select a SU $s$ to cooperate with the
PU at $i$th power level with a certain probability that depends on
$m$. 
\item When $Q_{p}(t)=0,$ select a SU $s$ to transmit its own packets at
$i$th power level with a certain probability. 
\end{itemize}
Consider a subclass of the policies in $\mathcal{C}_{1}$, denoted
by $\mathcal{C}_{0}$, which consists of policies whose decisions
are based solely on whether the PU queue is zero or not. In each time
slot $t$, a policy in ${\cal C}_{0}$ acts as follows: 
\begin{itemize}
\item When $Q_{p}(t)>0,$ or equivalently the PU channel is sensed busy,
select a SU $s$ to cooperate at $i$th power level with a probability
$q\left(s,i\left|b\right.\right)$. 
\item When $Q_{p}(t)=0,$ or equivalently the PU channel is sensed idle,
select a SU $s$ to transmit its own data at $i$th power level with
probability $q\left(s,i\left|e\right.\right)$. 
\end{itemize}
Since the policies in ${\cal C}_{0}$ are not based on the actual
value of $Q_{p}\left(t\right)$, but only whether $Q_{p}\left(t\right)$
is greater than or equal to zero, it follows that ${\cal C}_{0}\subseteq{\cal C}_{1}$. 

For the analysis that follows, it is helpful to introduce the extended
class of policies $\mathcal{C}_{2}$ which follow the policy constraints
with the exception the PU priority constraint, i.e., when the PU queue
is non-empty at the beginning of a slot, the policy may select to
transmit one of the SU packets instead of a PU packet. In this case,
the available controls at the beginning of each time slot are of the
form $(u,s,i),\ u\in\left\{ 1,0\right\} ,\ s\in\mathcal{S},\ i\in\mathcal{I}_{s}^{0},$
where 
\begin{itemize}
\item Control $\left(1,s,i\right)$, dictates transmission of PU traffic
and assigns SU $s$ at $i$th power level to cooperate with the PU.
Note that this control can be assigned even if the PU queue is empty,
in which case no packet is transmitted. 
\item Control $\left(0,s,i\right)$, dictates transmission of only SU traffic,
and selects SU $s$ to transmit at $i$th power level. 
\end{itemize}
Since policies in $\mathcal{C}_{2}$ do not impose the PU priority
constraint, and they may include even non-stationary policies, it
follows that ${\cal {\cal C}}_{1}\subseteq\mathcal{C}_{2}.$ Hence,
it holds that ${\cal C}_{0}\subseteq{\cal C}_{1}\subseteq{\cal C}_{2}$
and the corresponding achievable rate regions $\mathcal{R}_{0},\ \mathcal{R}_{1},\ \mathcal{R}_{2}$,
satisfying the policy constraints under the classes of policies $\mathcal{C}_{0},\ {\cal {\cal C}}_{1},\ C_{2}$,
satisfy $\mathcal{R}_{0}\subseteq{\cal {\cal R}}_{1}\subseteq\mathcal{R}_{2}.$

\begin{figure}
\centering\includegraphics[width=0.54\columnwidth]{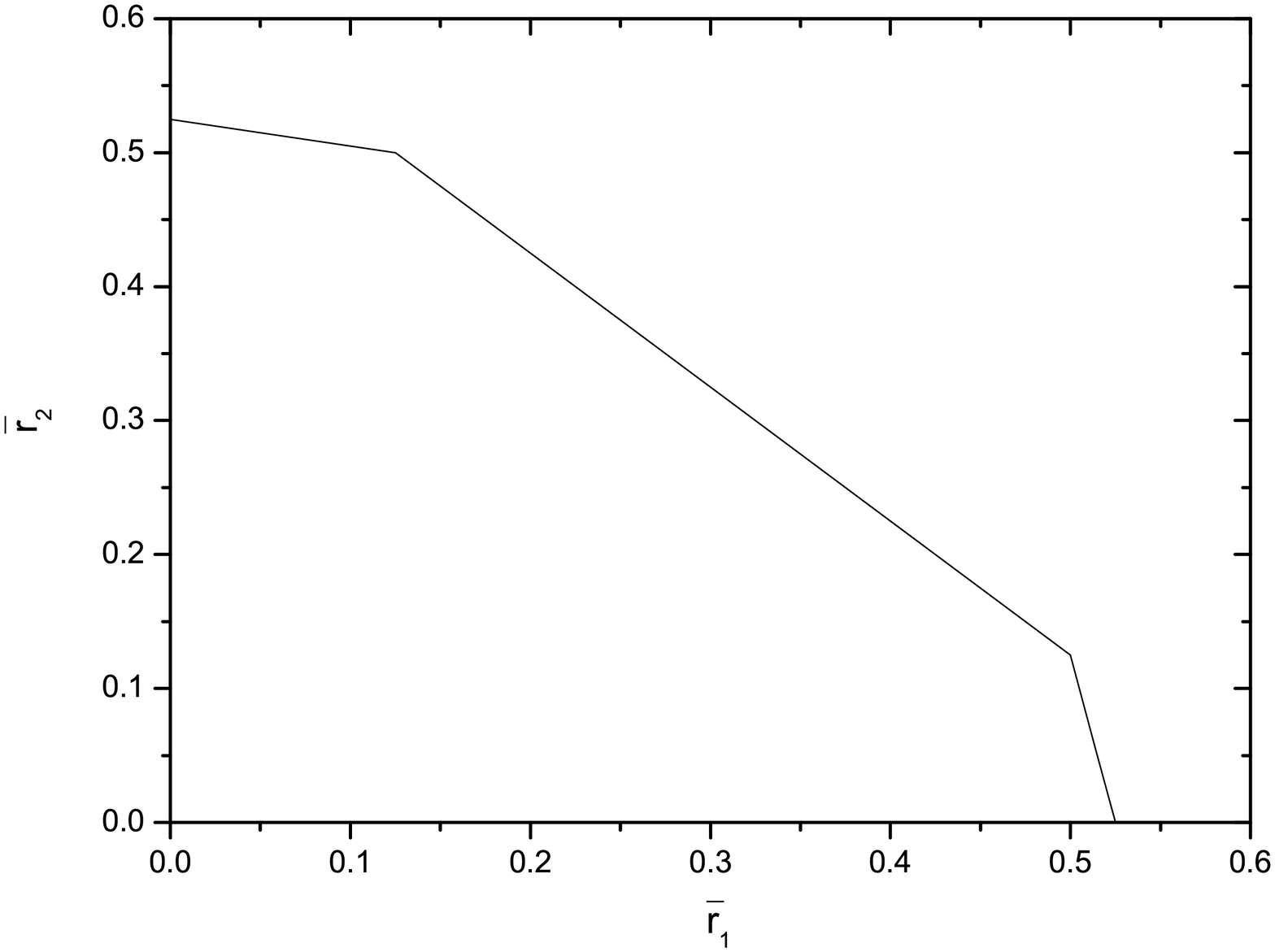}

\protect\caption{The rate regions ${\cal R}_{0}$, ${\cal R}_{1}$ and ${\cal R}_{2}$,
which coincide, for the system setup scenario with ${\cal S}=\left\{ 1,2\right\} $,
$\lambda_{p}=0.3,$ and $\mathcal{I}_{s}^{0}=\left\{ 0,1,2,3,4\right\} $,
$\mathbf{P}_{s}=\left\{ 0,0.25,0.5,0.75,1\right\} $, $r_{p}\left(0\right)=0.4$,
$r_{p}\left(s,1\right)=0.5$, $r_{p}\left(s,2\right)=0.6$, $r_{p}\left(s,3\right)=0.7$,
$r_{p}\left(s,4\right)=0.8$, $r_{s}\left(1\right)=0.3$, $r_{s}\left(2\right)=0.5$,
$r_{s}\left(3\right)=0.8$, $r_{s}\left(4\right)=1$, $\hat{P}_{s}=0.5$,
for all $s\in\mathcal{S}$.}
\label{Fig:region}
\end{figure}

It might seem at first glance that a policy in class ${\cal C}_{0}$
with a restricted control space will lead to suboptimal performance.
However, this is not the case. In the next section we show that $\mathcal{R}_{2}\subseteq\mathcal{R}_{0}$,
thus reaching the interesting key conclusion that $\mathcal{R}_{0}={\cal {\cal R}}_{1}=\mathcal{R}_{2}.$
The rate regions ${\cal R}_{0},$ ${\cal R}_{1}$ and ${\cal R}_{2}$
(which coincide) for a particular system setup scenario with 2 SUs
are illustrated in Fig. \ref{Fig:region}. Hence, under any optimization
objective, \emph{it suffices to restrict attention to policies in
$\mathcal{C}_{0}$ even if one has the freedom of not adhering to
the PU priority constraint.}

\section{\label{sec:Prformance-regions-identical}Characterization of Achievable
Rate Regions $\mathcal{R}_{0}$, $\mathcal{R}_{1}$, $\mathcal{R}_{2}$}

In this section we substantiate our previous claim. Towards this end,
we first determine the achievable rate region of policies in $\mathcal{C}_{0}$,
namely $\mathcal{R}_{0}$, in subsection (\ref{sub:Performance-Region-C_0}),
as well as the stability region of the PU queue when policies in class
$\mathcal{C}_{0}$ are employed. Second, we determine the achievable
rate region of policies in $\mathcal{C}_{2}$, namely $\mathcal{R}_{2}$,
in subsection (\ref{sub:Performance-Region-C_2}), and finally we
prove that $\mathcal{R}_{0}$ coincides with $\mathcal{R}_{2}$.

\subsection{\label{sub:Performance-Region-C_0}Achievable Rate Region of Policies
in Class $\mathcal{C}_{0}$}

For a given policy $\pi$ in class ${\cal C}_{0}$, the average packet
service rate of the PU queue is given by 
\begin{equation}
\bar{r}_{p}=\sum_{s\in\mathcal{S}}\sum_{i\in\mathcal{I}_{s}^{0}}r_{p}(s,i)q(s,i\left|b\right.).\label{eq:aver_prim_rate}
\end{equation}
Standard results from queuing theory show that the stability region
of the PU queue under $\pi$, that is, the closure of the set of PU
arrival rates $\lambda_{p}$ for which the PU queue is mean-rate stable
\cite{B:Neely_book}, is the set of arrival rates that fall in the
interval $\left[0,\bar{r}_{p}\right]$. Assume next that $\lambda_{p}\in\left[0,\bar{r}_{p}\right)$
(so that the PU queue is stable) and let $q_{b}$ be the steady state
probability that the PU queue is busy under $\pi$. Viewing the transmitter
at the PU as a queuing system holding 0 (if the PU queue is empty)
or 1 packets (i.e., the packet whose transmission is attempted if
the PU queue is non-empty) and applying Little's formula to this system,
we have 
\begin{equation}
q_{b}=\Pr\left\{ \text{PU queue is non-empty}\right\} =\frac{\lambda_{p}}{\bar{r}_{p}}.\label{eq:little_theorem}
\end{equation}
Hence, the steady state probability that the PU queue is empty is
$q_{e}=1-q_{b}.$ Due to the imposed PU priority constraint, SUs may
transmit their own data only when the PU queue is empty. Hence, the
average packet transmission rate of SU $s$ traffic is equal to 
\begin{align}
\bar{r}_{s} & =\left(\sum_{i\in{\cal I}_{s}}r_{s}\left(i\right)q\left(s,i\left|e\right.\right)\right)q_{e}.\label{eq:h1}
\end{align}

The average power consumption of SU $s\in\mathcal{S}$ is 
\begin{align}
\bar{P}_{s} & =q_{e}\sum_{i\in\mathcal{I}_{s}}P_{s}\left(i\right)q(s,i\left|e\right.)+q_{b}\sum_{i\in\mathcal{I}_{s}}P_{s}\left(i\right)q(s,i\left|b\right.)\label{eq:Aver-primary}
\end{align}
and since $\pi\in\mathcal{C}_{0},$ it satisfies the power constraints
(\ref{eq:power_constraint}), i.e., $\bar{P}_{s}\leq\hat{P}_{s},\, s\in\mathcal{S}$.
The discussion above shows that the constraints that need to be satisfied
by the set of probabilities $\left\{ q_{b},q\left(s,i\left|b\right.\right),q\left(s,i\left|e\right.\right),q_{e}\right\} \ s\in\mathcal{S}$,
according to (\ref{eq:power_constraint}), (\ref{eq:little_theorem}),
are given by 
\begin{equation}
q_{b}\sum_{s\in\mathcal{S}}\sum_{i\in\mathcal{I}_{s}^{0}}r_{p}(s,i)q(s,i\left|b\right.)=\lambda_{p}\label{eq:rate_equality}
\end{equation}
\begin{equation}
q_{e}\sum_{i\in\mathcal{I}_{s}}P_{s}\left(i\right)q(s,i\left|e\right.)+q_{b}\sum_{i\in\mathcal{I}_{s}}P_{s}\left(i\right)q(s,i\left|b\right.)\leq\hat{P}_{s},\ s\in\mathcal{S}\label{eq:aver_power}
\end{equation}
\begin{equation}
q_{b}+q_{e}=1\label{eq:sum_prob_busy}
\end{equation}
\begin{equation}
\sum_{s\in{\cal S}}\sum_{i\in{\cal I}_{s}^{0}}q\left(s,i\left|b\right.\right)=1\label{eq:sum_cond_prob_1}
\end{equation}
\begin{equation}
\sum_{s\in{\cal S}}\sum_{i\in{\cal I}_{s}^{0}}q\left(s,i\left|e\right.\right)=1\label{eq:sum_cond_prob_2}
\end{equation}
\begin{equation}
q_{b}\geq0,\, q_{e}\geq0,\, q\left(s,i\left|b\right.\right)\geq0,\, q\left(s,i\left|e\right.\right)\geq0,\ \ s\in\mathcal{S},\ i\in\mathcal{I}_{s}^{0}\label{eq:non_negativity}
\end{equation}

Conversely, given the set of probabilities $\left\{ q_{b},q\left(s,i\left|b\right.\right),q\left(s,i\left|e\right.\right),q_{e}\right\} _{s\in\mathcal{S},\ i\in\mathcal{I}_{s}^{0}}$
that satisfy the constraints (\ref{eq:rate_equality})-(\ref{eq:non_negativity}),
with $q_{b}<1\mbox{,}$ an admissible policy in $\mathcal{C}_{0}$
can be defined. Hence, the performance space of these policies is
the set of $\mathbf{\bar{r}}$ defined by (\ref{eq:h1}), where the
set of probabilities $\left\{ q_{b},q\left(s,i\left|b\right.\right),q\left(s,i\left|e\right.\right),q_{e}\right\} _{s\in\mathcal{S},\ i\in\mathcal{I}_{s}^{0}}$
satisfy constraints (\ref{eq:rate_equality})-(\ref{eq:non_negativity}).

While constraints of (\ref{eq:rate_equality})-(\ref{eq:non_negativity})
are nonlinear with respect to parameters $\left\{ q_{b},q\left(s,i\left|b\right.\right),q\left(s,i\left|e\right.\right),q_{e}\right\} $,
they can be easily transformed into linear ones through the transformation
\begin{equation}
q\left(b,s,i\right)=q_{b}q\left(s,i\left|b\right.\right),\, q\left(e,s,i\right)=q_{e}q\left(s,i\left|e\right.\right).\label{eq:lineartransformation}
\end{equation}
Note that $q\left(b,s,i\right)$ is the probability that the PU is
busy \emph{and} SU $s$ is selected for cooperation at power level
$i$, while $q\left(e,s,i\right)$ is the probability that the PU
is idle \emph{and }SU $s$ packets are transmitted in a slot at power
level $i.$ With this transformation, the constraints that characterize
the achievable rate region of policies in $\mathcal{C}_{0}$ become,
\begin{equation}
\sum_{s\in{\cal S}}\sum_{i\in{\cal I}_{s}^{0}}r_{p}\left(s,i\right)q\left(b,s,i\right)=\lambda_{p}\label{eq:rate_prim_constr_3}
\end{equation}
\begin{equation}
\sum_{i\in{\cal I}_{s}}P_{s}\left(i\right)q\left(e,s,i\right)+\sum_{i\in{\cal I}_{s}}P_{s}\left(i\right)q\left(b,s,i\right)\leq\hat{P}_{s},\ s\in\mathcal{S}\label{eq:power_aver_constr_3}
\end{equation}

\begin{equation}
\sum_{s\in{\cal S}}\sum_{i\in{\cal I}_{s}^{0}}q\left(e,s,i\right)+\sum_{s\in{\cal S}}\sum_{i\in{\cal I}_{s}^{0}}q\left(b,s,i\right)=1\label{eq:sum_of_prob_3}
\end{equation}

\begin{equation}
q\left(e,s,i\right)\geq0\ q(b,s,i)\geq0,\ s\in\mathcal{S},\ i\in\mathcal{I}_{s}^{0}.\label{eq:nonnegative1}
\end{equation}
In addition, the achievable rate of each SU $s\in\mathcal{S}$, given
by (\ref{eq:h1}), can be rewritten as 
\begin{equation}
\bar{r}_{s}=\sum_{i\in{\cal I}_{s}}r_{s}\left(i\right)q\left(e,s,i\right)\label{eq:aver_second1}
\end{equation}
In fact, it can shown that (\ref{eq:h1}) and (\ref{eq:rate_equality})-(\ref{eq:non_negativity}),
define the same performance space as (\ref{eq:rate_prim_constr_3})-(\ref{eq:aver_second1}).
This is described in the following proposition. 
\begin{prop}
\label{thm:Equivalence}The performance space of $\left\{ \bar{r}_{s}\right\} $
which is defined by Eqs. (\ref{eq:h1}) and (\ref{eq:rate_equality})-(\ref{eq:non_negativity})
is equivalent with the corresponding performance space defined by
Eqs. (\ref{eq:rate_prim_constr_3})-(\ref{eq:aver_second1}). \end{prop}
\begin{IEEEproof}
Please refer to Appendix \ref{sec:Proof-of-Equivalence}. 
\end{IEEEproof}
In the next section, we use the characterization of the achievable
rate region of policies in $\mathcal{C}_{0}$ in terms of constraints
(\ref{eq:rate_prim_constr_3})-(\ref{eq:aver_second1}) to show that
this region coincides with the achievable rate region of policies
in $\mathcal{C}_{2}$.

\subsubsection{Stability region of PU Queue under the class of policies in $\mathcal{C}_{0}$}

Based on the discussion above, the stability region of the PU queue
under the class of policies in $\mathcal{C}_{0}$ is the set of $\lambda_{p}$
for which there exists a set of probabilities $\left\{ q\left(b,s,i\right),\ q\left(e,s,i\right)\right\} _{s\in\mathcal{S},\ i\in\mathcal{I}_{s}^{0}}$
that satisfy (\ref{eq:rate_prim_constr_3})-(\ref{eq:aver_second1}).
Based on this observation we have the following corollary. 
\begin{cor}
\label{cor:The-stability-region-1}The stability region of the PU
queue under the class of policies in $\mathcal{C}_{0}$ is the interval
$[0,\hat{\lambda}]$ where $\hat{\lambda}$ is the resulting value
of the objective of the following linear optimization problem in terms
of $x\left(b,s,i\right)$, for all $s\in\mathcal{S}$ and $i\in\mathcal{I}_{s}^{0}$.
\begin{eqnarray}
\text{maximize:} & \sum_{s\in\mathcal{S}}\sum_{i\in\mathcal{I}_{s}^{0}}r_{p}(s,i)x(b,s,i)\label{eq:max_rate_equality-1}\\
\text{subject to} & \sum_{i\in\mathcal{I}_{s}}P_{s}\left(i\right)x(b,s,i)\leq\hat{P}_{s},\ s\in\mathcal{S}\label{eq:max_aver_power-1}\\
 & \sum_{s\in{\cal S}}\sum_{i\in{\cal I}_{s}^{0}}x\left(b,s,i\right)\leq1\label{eq:max_sum_cond_prob-1}\\
 & x\left(b,s,i\right)\geq0,\ s\in\mathcal{S},\ i\in\mathcal{I}_{s}^{0}\label{eq:nonnegative3-1}
\end{eqnarray}
\end{cor}
\begin{IEEEproof}
Please refer to Appendix \ref{sec:Proof-of-max_rate}.\end{IEEEproof}
\begin{rem}
It can be easily seen that the value of optimization problem in Corollary
\ref{cor:The-stability-region-1} does not change if inequality in
(\ref{eq:max_sum_cond_prob-1}) is replaced by equality. This implies
what is intuitively expected, i.e., when $\lambda_{p}=\hat{\lambda}$,
no idle slots are left by PU, i.e., $q_{b}=1$ and $q_{e}=0$, and
\emph{the available power from any SU is allocated only to the cooperation
with the PU}. 
\end{rem}

\subsubsection{Implementation of policies in class $\mathcal{C}_{0}$}

In order to implement the policies in the proposed restricted class
$\mathcal{C}_{0}$, the probabilities $\left\{ q\left(e,s,i\right),q\left(b,s,i\right)\right\} _{i\in\mathcal{I}_{s}^{0},\: s\in\mathcal{S}}$
need to be determined. These probabilities are obtained through solving
the following optimization problem OPT0

\begin{eqnarray}
\textrm{maximize} & f\left(\mathbf{\bar{r}}\right)\label{eq:Mainproblemobjective}\\
\textrm{subject to} & \sum_{s\in{\cal S}}\sum_{i\in{\cal I}_{s}^{0}}r_{p}\left(s,i\right)q\left(b,s,i\right)=\lambda_{p}\label{eq:Mainproblemconstr1}\\
 & \sum_{i\in{\cal I}_{s}}P_{s}\left(i\right)q\left(e,s,i\right)+\sum_{i\in{\cal I}_{s}}P_{s}\left(i\right)q\left(b,s,i\right)\leq\hat{P}_{s},\ s\in\mathcal{S}\label{eq:Mainproblemconstr2}\\
 & \sum_{s\in{\cal S}}\sum_{i\in{\cal I}_{s}^{0}}q\left(e,s,i\right)+\sum_{s\in{\cal S}}\sum_{i\in{\cal I}_{s}^{0}}q\left(b,s,i\right)=1\label{eq:Mainproblemconstr3}\\
 & q\left(e,s,i\right)\geq0\ q(b,s,i)\geq0,\ s\in\mathcal{S},\ i\in\mathcal{I}_{s}^{0},\label{eq:Mainproblemconstr4}
\end{eqnarray}
where $\mathbf{\bar{r}\triangleq}\left\{ \bar{r}_{s}\right\} _{s\in\mathcal{S}}$,
and $\bar{r}_{s}=\sum_{i\in{\cal I}_{s}}r_{s}\left(i\right)q\left(e,s,i\right)$.
In problem OPT0 the optimization variables are $\left\{ q\left(e,s,i\right),q\left(b,s,i\right)\right\} _{i\in\mathcal{I}_{s}^{0},\: s\in\mathcal{S}}$,
whereas $r_{p}\left(s,i\right)$, $r_{s}\left(i\right)$, $P_{s}\left(i\right)$,
for all $i\in\mathcal{I}_{s}^{0}$, and $s\in\mathcal{S}$, are fixed
system model parameters. Specifically, $r_{p}\left(s,i\right)$ denotes
the probability of successful transmission of the PU packet when SU
$s$ cooperates at $i$th power level, while $r_{s}\left(i\right)$
denotes the probabilty of successful transmission of SU $s$ packet,
when SU $s$ transmits at $i$th power level. $P_{s}\left(i\right)$
denotes the transmit power that corresponds to level $i\in\mathcal{I}_{s}^{0}$
that SU $s$ uses in either case, and $\hat{P}_{s}$ denotes the maximum
average transmit power available for SU $s$. Constraint (\ref{eq:Mainproblemconstr1})
ensures that the average packet service rate of the PU queue equals
its average input rate, $\lambda_{p}$, and, therefore, guarantees
stability of the PU queue. The inequality constraints in (\ref{eq:Mainproblemconstr2})
are the long-term average power constraints for all SUs. Finally,
constraints (\ref{eq:Mainproblemconstr3}) and (\ref{eq:Mainproblemconstr4})
are imposed because the optimization variables $\left\{ q\left(e,s,i\right),q\left(b,s,i\right)\right\} _{i\in\mathcal{I}_{s}^{0},\: s\in\mathcal{S}}$
represent probabilities. In case where the selected objective function
in (\ref{eq:Mainproblemobjective}), $f\left(\cdot\right)$, is a
concave function of $\bar{\mathbf{r}},$ then, problem OPT0 is a convex
optimization problem which can be solved efficiently via interior
point methods. Once variables $\left\{ q\left(e,s,i\right),q\left(b,s,i\right)\right\} _{i\in\mathcal{I}_{s}^{0},\: s\in\mathcal{S}}$
are determined, we can obtain the probabilities $\left\{ q_{b},q\left(s,i\left|b\right.\right),q\left(s,i\left|e\right.\right),q_{e}\right\} _{s\in\mathcal{S},\ i\in\mathcal{I}_{s}^{0}}$
through the linear transformation in (\ref{eq:lineartransformation}).
Then, policies in $\mathcal{C}_{0}$ act as we describe in section
\ref{sub:Control-Objective}.

\subsection{\label{sub:Performance-Region-C_2}Achievable Rate Region of Policies
in Class $\mathcal{C}_{2}$}

Contrary to the available controls when the PU priority constraint
is imposed, the set of available controls for policies in $\mathcal{C}_{2}$
does not obey the PU priority constraint (thus, a slot may be allocated
to SU packet transmission, even if the PU queue is nonempty). Hence,
this class of policies falls in the framework of policies studied
in \cite{B:Neely_book}, whose achievable rate region can be characterized
again by the achievable rate region of stationary policies. In the
latter framework, a stationary policy selects at the beginning of
each time slot the control $\left(u,s,i\right)$ with probability
$p\left(u,s,i\right)$. Under such a policy, the probability of successful
transmission of SU $s$ packets is 
\begin{equation}
\bar{r}_{s}=\sum_{i\in{\cal I}_{s}}r_{s}\left(i\right)p\left(0,s,i\right),\label{eq:SecondaryRate}
\end{equation}
while, the probability of successful transmission of PU packets is
\begin{equation}
\bar{r}_{p}=\sum_{s\in\mathcal{S}}\sum_{i\in\mathcal{I}_{s}^{0}}r_{p}(s,i)p(1,s,i),\label{eq:PrimaryRate}
\end{equation}
and stability of the PU queue requires that 
\begin{equation}
\bar{r}_{p}\geq\lambda_{p}.\label{eq:rate_equality-2}
\end{equation}
Also, the average power constraint requirement implies that 
\begin{equation}
\sum_{i\in\mathcal{I}_{s}}P_{s}\left(i\right)p(0,s,i)+\sum_{i\in\mathcal{I}_{s}}P_{s}\left(i\right)p(1,s,i)\leq\hat{P}_{s},\, s\in\mathcal{S}.\label{eq:aver_power-2}
\end{equation}
Finally, since $p\left(u,s,i\right)$ are probabilities, we must have
\begin{align}
\sum_{s\in{\cal S}}\sum_{i\in{\cal I}_{s}^{0}}p\left(0,s,i\right)+\sum_{s\in{\cal S}}\sum_{i\in{\cal I}_{s}^{0}}p\left(1,s,i\right) & =1\label{eq:sum_prob_busy-2}\\
p(0,s,i)\geq0,\ p(1,s,i)\geq0,\text{\,\ s\ensuremath{\in\mathcal{S}},} & \, i\in\mathcal{I}_{s}^{0}.\label{eq:nonnetative2}
\end{align}

Constraints (\ref{eq:rate_equality-2})-(\ref{eq:nonnetative2}) together
with (\ref{eq:SecondaryRate}) define the achievable rate region ${\cal R}_{2}$
of policies in $\mathcal{C}_{2}$. The similarity of these constraints
compared to those in (\ref{eq:rate_prim_constr_3})-(\ref{eq:aver_second1})
should be noted. From a math perspective, the only difference is that
there exists equality in (\ref{eq:rate_prim_constr_3}), as opposed
to inequality in (\ref{eq:rate_equality-2}). However, there is difference
in the interpretation of these probabilities. Specifically, 
\begin{itemize}
\item $q\left(b,s,i\right)$ is the probability that PU queue is nonempty
\emph{and} SU $s$ is selected for cooperation at $i$th power level,
while $p(1,s,i)$ is the probability that SU $s$ is selected for
cooperation at $i$th power level and dictating PU transmission as
well (irrespective of the PU queue size). 
\item $q\left(e,s,i\right)$ is the probability that PU queue is empty \emph{and
}secondary user $s$ packets are transmitted in a slot at $i$th power
level, while $p(0,s,i)$ is the probability of selecting secondary
user $s$ packet for transmission at the $i$th power level, while
PU does not transmit (irrespective of the PU queue size). 
\end{itemize}
As discussed earlier, since ${\cal C}_{0}\subseteq{\cal C}_{2}$,
${\cal R}_{0}\subseteq{\cal R}_{2}$. The next theorem shows that
${\cal R}_{2}={\cal R}_{0}$. 
\begin{thm}
\label{thm:Equivalence2}It holds ${\cal R}_{2}\subseteq{\cal R}_{0}$,
hence $\mathcal{R}_{0}={\cal {\cal R}}_{1}=\mathcal{R}_{2}.$\end{thm}
\begin{IEEEproof}
Please refer to the Appendix \ref{sec:Proof-of-Equivalence2}. 
\end{IEEEproof}

\section{Extensions to the basic model\label{sec:Extensions}}

In this section, we extend the model that has been investigated so
far in two directions. First, we assume exogenous packet arrivals
to the SU queues, instead of infinite queue backlogs. Second, imperfect
channel sensing effects are taken into account.

\subsection{Incorporating Exogenous Packet arrivals to SU queues \label{sec:Incorporating-Exogenous-Packet}}

In this part, we investigate the scenario where packets arrive exogenously
to SU queues. Specifically, we assume that at the beginning of slot
$t,$ $A_{s}(t)$ packets arrive to the queue of SU. Furthermore,
for a given SU $s$, $A_{s}(t),\ t=0,1...$ are i.i.d random variables
with $\mathbb{E}\left[A_{s}(t)\right]=\lambda_{s},\ \mathbb{E}\left[\left(A_{s}\left(t\right)\right)^{2}\right]<\infty$
and the arrival processes $\left\{ A_{s}(t)\right\} _{t=0}^{\infty},\ s\in{\cal S}$
are independent of each other. Regarding the packet arrival process
to the PU queue, $A_{p}\left(t\right)$, we also assume that it consists
of i.i.d. random variables and is independent of the arrival processes
to the SU queues.

\subsubsection{Admissible Policies }

As in the case where the SU queues were backlogged, an admissible
policy should satisfy the constraints\textit{ }described in section
\ref{sub:Control-Objective}\textit{. }Regarding SU queues, there
are no constraints on the rates of their arrival processes. Hence,
depending on the arrival rates to these queues, they may be stable
or unstable. To deal with the issue of instability, we assume that
flow control is applied to each of the SU queues, which has the following
form \cite{B:Neely_book}: among the $A_{s}(t)$ packets that arrive
at the queue of SU $s,$ a number $B_{s}(t)\leq A_{s}(t)$ is accepted
by the system and the rest (if any) are dropped. Thus, the \emph{flow
control objective} is that the SU queues with input the $B_{s}(t)$
packets must be mean rate stable.

In general, the admissible policies in this setup take control actions
at time slot $t$, based on the history of the system up to time $t,$
which includes queue sizes of the PU and SU queues up to time $t$.
We call this class of policies $\tilde{{\cal C}}_{1}$. Similar to
the previous analysis, we consider a subclass of policies in $\tilde{{\cal C}}_{1}$,
denoted by $\tilde{{\cal C}}_{0}$, which consists of policies whose
decisions are based solely on whether the PU queue is empty or not,
hence not requiring information about the queue sizes at the PU and
SU queues. In each time slot $t$, a policy in $\tilde{{\cal C}}_{0}$
acts as follows: 
\begin{itemize}
\item \emph{Flow control action}: Each of the $A_{s}(t)$ packets that arrive
to SU $s$ at time $t,$ is admitted with probability $p_{s}^{a}$.
The packet admission events are independent of each other and independent
of other processes in the system. 
\item When $Q_{p}(t)>0$, select a SU $s$ to cooperate at $i$th power
level with a probability $q\left(s,i\left|b\right.\right)$. 
\item When $Q_{p}(t)=0$, select a SU $s$ to transmit its own data at $i$th
power level with probability $q\left(s,i\left|e\right.\right)$. If
the selected SU has no data to transmit, it loses its transmission
opportunity. 
\end{itemize}
For performance comparison, we consider the extended class of policies
$\tilde{{\cal C}}_{2}$ which employs flow control at the SU queues
and obeys all constraints of policies in $\tilde{C}_{1}$, except
the PU priority constraint. Hence we again have $\tilde{{\cal C}}_{0}\subseteq\tilde{{\cal C}}_{1}\subseteq\tilde{{\cal C}}_{2}$.
The performance measure of interest in this case is the throughput
of SU queues, i.e., the long term average number of packets per slot,
$R_{s},$ that are delivered to the receiver of SU $s,\ s\in{\cal S}$.
The set of achievable throughput vectors $\boldsymbol{R}=\left\{ R_{s}\right\} _{s\in{\cal S}}$
under class of policies $\tilde{{\cal C}_{i}},\ i=0,1,2$, is denoted
by $\tilde{{\cal R}}_{i}$. Since $\tilde{{\cal C}}_{0}\subseteq\tilde{{\cal C}}_{1}\subseteq\tilde{{\cal C}}_{2}$
we again have, $\tilde{{\cal R}}_{0}\subseteq\tilde{{\cal R}}_{1}\subseteq\tilde{{\cal R}}_{2}.$

\subsubsection{Throughput Regions of Policies in Classes ${\cal \tilde{C}}_{0}$
and ${\cal \tilde{C}}_{2}$}

Similarly to the analysis in Section \ref{sub:Performance-Region-C_0},
it can be shown that $\tilde{{\cal R}}_{0}$ consists of all vectors
$\boldsymbol{R}=\left\{ R_{s}\right\} _{s\in{\cal S}}$ that satisfy
\begin{equation}
R_{s}\leq\min\left\{ \lambda_{s},\bar{r}_{s}\right\} ,\ s\in{\cal S}\label{mod_rate_SU}
\end{equation}
where $\bar{r}_{s}$ is defined by (\ref{eq:rate_prim_constr_3})-(\ref{eq:nonnegative1})
and (\ref{eq:aver_second1}). Note that in the current setup, $\bar{r}_{s}$
represents the ``offered'' service rate to SU $s$ queue, i.e.,
the probability of successful transmission of an SU $s$ packet. For
maximizing the throughput of each SU queue, we must have $R_{s}=\min\left\{ \lambda_{s},\bar{r}_{s}\right\} $.
Moreover, since flow control is chosen to stabilize the SU queues,
we must have $R_{s}=\lambda_{s}p_{s}^{a}$, with $p_{s}^{a}=\frac{\min\left\{ \lambda_{s},\bar{r}_{s}\right\} }{\lambda_{s}},\ s\in{\cal S}$.

On the other hand, for the stationary policies in $\tilde{{\cal C}}_{2}$,
it can be shown \cite{B:Neely_book} that $\tilde{{\cal R}}_{2}$
consists of all vectors that satisfy (\ref{mod_rate_SU}) and $\sum_{s\in{\cal S}}\sum_{i\in{\cal I}_{s}^{0}}r_{p}\left(s,i\right)q\left(b,s,i\right)\geq\lambda_{p},$
with $\bar{r}_{s}$ being defined by (\ref{eq:power_aver_constr_3})-(\ref{eq:nonnegative1})
and (\ref{eq:aver_second1}).

Based on the structure of the throughput regions described above,
it follows by a similar argument as in section \ref{sec:Prformance-regions-identical}
that $\tilde{{\cal R}}_{0}=\tilde{{\cal R}}_{2}$, which implies again
that policies in $\tilde{{\cal C}}_{0}$ can achieve any throughput
vector achievable by the less restrictive policies in $\tilde{{\cal C}}_{2}$.

\subsubsection{Selecting Optimal Policies in $\tilde{{\cal C}}_{0}$}

Consider the problem of selecting a policy in $\tilde{{\cal C}}_{0}$
that maximizes $f(\boldsymbol{R}),$ with $\boldsymbol{R}\in\tilde{{\cal R}}_{0}.$
Based on the above, it is then easy to see that this optimization
problem is equivalent to 
\begin{equation}
\text{maximize}\ f\left(\left\{ \min\left(\lambda_{s},\bar{r}_{s}\right)\right\} _{s\in{\cal S}}\right),\label{eq:throughputOptimize}
\end{equation}
where $\bar{r}_{s}$ is defined by (\ref{eq:aver_second1}) and (\ref{eq:rate_prim_constr_3})-(\ref{eq:nonnegative1}).

\subsection{Imperfect Sensing \label{sec:Imperfect-Sensing}}

In this part, we investigate the effects of imperfect sensing on the
mode of operation and the performance of policies in ${\cal C}_{0}$.
For simplicity we assume that the SUs are infinitely backlogged. The
case where packets arrive randomly at the SUs can be handled in a
similar fashion as in section \ref{sec:Incorporating-Exogenous-Packet}.

We assume that cooperative sensing takes place, so that all SUs make
the same decision at each slot as to whether the primary channel is
busy or idle. We assume that PU channel sensing events are independent
across slots and independent of the transmission choices of the users.
We denote the probabilities of detection and false alarm of the sensing
mechanism as ${\cal P}_{D}=\Pr\left\{ \text{sense busy\ensuremath{\left|\text{channel is busy}\right.}}\right\} $
and ${\cal P}_{F}=\Pr\left\{ \text{sense busy\ensuremath{\left|\text{channel is idle}\right.}}\right\} $,
respectively. Two sources of error and inefficiency may occur in this
situation: 
\begin{itemize}
\item The primary channel is busy but sensed idle (an event occurring with
probability $1-\mathcal{P}_{D}$). We distinguish two subcases:

\begin{itemize}
\item One of the SUs transmits its own packet at the same slot with the
PU, an event with probability $1-\sum_{s\in\mathcal{S}}q\left(s,0\left|e\right.\right)$%
\footnote{Recall using power level $0$ implies no transmission.%
}. In this case, collision occurs and both transmissions fail. 
\item No SU transmits a packet, an event with probability $\sum_{s\in\mathcal{S}}q\left(s,0\left|e\right.\right)$.
In this case the PU transmission is successful with probability $r_{p}\left(0\right).$ 
\end{itemize}

The effect of this error on the probability of successful transmission
of PU packet is given by 
\begin{equation}
\bar{r}_{p}=\left(1-\mathcal{P}_{D}\right)\sum\limits _{s\in\mathcal{S}}q\left(s,0\left|e\right.\right)r_{p}\left(0\right)+\mathcal{P}_{D}\sum\limits _{s\in\mathcal{S}}\sum\limits _{i\in\mathcal{I}_{s}^{0}}r_{p}\left(s,i\right)q\left(s,i\left|b\right.\right)\label{eq:mod_r_p}
\end{equation}

\item When the PU channel is idle but it is sensed busy, an SU may be allocated
for cooperation with the PU, thus losing the opportunity to transmit
its own data. Hence, the probability of successful transmission of
SU packets is affected by the probability of the event that the PU
channel is idle and sensed idle (equal to $q_{e}\left(1-\mathcal{P}_{F}\right)$).
For the SU $s$, this probability becomes 
\begin{equation}
\bar{r}_{s}=q_{e}\left(1-\mathcal{P}_{F}\right)\sum_{i\in\mathcal{I}_{s}^{0}}r_{s}\left(s,i\right)q\left(s,i\left|e\right.\right).\label{eq:SenseSUrate}
\end{equation}

\end{itemize}
Regarding the average power consumed by SU $s$ under a policy in
$\mathcal{C}_{0}$, we consider the following events: 
\begin{enumerate}
\item The event that PU channel is busy and is sensed busy, with probability
$q_{b}\mathcal{P}_{D}.$ Then, SU $s$ consumes an average power of
$\sum_{i\in\mathcal{I}_{s}^{0}}P_{s}\left(i\right)q\left(s,i\left|b\right.\right).$
\item The event that PU channel is busy and is sensed idle, with probability
$q_{b}\left(1-\mathcal{P}_{D}\right).$ Then, SU $s$ consumes an
average power of $\sum_{i\in\mathcal{I}_{s}^{0}}P_{s}\left(i\right)q\left(s,i\left|e\right.\right).$
\item The event that PU channel is idle and is sensed idle, with probability
$q_{e}\left(1-\mathcal{P}_{F}\right)$. Then, SU $s$ consumes an
average power of $\sum_{i\in\mathcal{I}_{s}^{0}}P_{s}\left(i\right)q\left(s,i\left|e\right.\right).$
\item The event that PU channel is idle and is sensed busy, with probability
$q_{e}\mathcal{P}_{F}$. Then, SU $s$ consumes an average power of
$\sum_{i\in\mathcal{I}_{s}^{0}}P_{s}\left(i\right)q\left(s,i\left|b\right.\right).$
\end{enumerate}
Based on the above, the new performance space when channel sensing
errors are introduced is determined by (\ref{eq:sum_prob_busy})-(\ref{eq:non_negativity})
and 
\begin{equation}
q_{b}\mathcal{P}_{D}\sum_{s\in\mathcal{S}}\sum_{i\in\mathcal{I}_{s}^{0}}r_{p}\left(s,i\right)q\left(s,i\left|b\right.\right)+q_{b}\left(1-\mathcal{P}_{D}\right)r_{p}\left(0\right)\sum_{s\in\mathcal{S}}q\left(s,0\left|e\right.\right)=\lambda_{p}.\label{eq:mod_Lit}
\end{equation}
\begin{equation}
\left(q_{b}\mathcal{P}_{D}+q_{e}\mathcal{P}_{F}\right)\sum_{i\in\mathcal{I}_{s}^{0}}P_{s}\left(i\right)q\left(s,i\left|b\right.\right)+\left(1-q_{b}\mathcal{P}_{D}-q_{e}\mathcal{P}_{F}\right)\sum_{i\in\mathcal{I}_{s}^{0}}P_{s}\left(i\right)q\left(s,i\left|e\right.\right)\leq\hat{P}_{s},\ s\in\mathcal{S}.\label{eq:mod_pow_constr}
\end{equation}

We seek transmission policies that achieve the following objective,
OPT1: 
\begin{eqnarray}
\text{maximize} & f\left(\bar{r}_{s}\right)\label{OPt1}\\
\text{subject to} & \text{(\ref{eq:sum_prob_busy})-(\ref{eq:non_negativity}), (\ref{eq:mod_Lit})-(\ref{eq:mod_pow_constr})}
\end{eqnarray}
where $\bar{r}_{s}$ are given by (\ref{eq:SenseSUrate}).

Due to (\ref{eq:mod_Lit})-(\ref{eq:mod_pow_constr}), OPT1 is a non-convex
optimization problem and therefore it is difficult to be solved optimally.
One way to solve OPT1 numerically, is to fix $q_{b}$, in which case
the constraints become linear and the problem can be easily solved.
Let $g(q_{b})$ be the maximum value of the objective of OPT1 for
$q_{b}\in[0,1]$ (for some values of $q_{b}$ the problem may be infeasible).
We can then solve the one-dimensional problem: 
\begin{equation}
\mbox{maximize}\ g(q_{b})\label{eq:one_dim_problem}
\end{equation}
where $0\leq q_{b}\leq1$ and the maximum can be specified through
exhaustive linear search methods. However, based on the following
remark, we can restrict the region of possible $q_{b}$ values, where
linear search is performed. 
\begin{prop}
\label{Rem:Regio}The probability of PU being busy when imperfect
sensing takes place, varies within 
\begin{equation}
\frac{\lambda_{p}}{{\cal P}_{D}r_{p,max}+\left(1-{\cal P}_{D}\right)r_{p}\left(0\right)}\leq q_{b}\leq\min\left\{ \frac{\lambda_{p}}{{\cal P}_{D}r_{p}\left(0\right)},1\right\} \label{eq:space}
\end{equation}
where $r_{p,max}=\max_{s,i}\left\{ r_{p}(s,i)\right\} $.\end{prop}
\begin{IEEEproof}
The proof follows straightforwardly based on (\ref{eq:mod_Lit}) and
is given in Appendix \ref{sec:Proof-of-Remark}. 
\end{IEEEproof}
Solving the one-dimensional problem (\ref{eq:one_dim_problem}) by
exhaustive search may be computationally expensive. As will be seen
in section \ref{sec:Simulation-Results}, a large number of numerical
investigations suggest that $g(q_{b})$ is a concave function of $q_{b}$.
We have not been able to prove rigorously that this property holds.
However, if it is indeed true, binary search methods can be used instead
for the solution of (\ref{eq:one_dim_problem}), thus reducing the
computational complexity from $\mathbf{\mathcal{M}}$ to $\log_{2}\mathcal{M}$,
where $\mathcal{M}$ stands for the number of values of $q_{b}$ investigated
in the space given by (\ref{eq:space}).

\section{Distributed Implementation \label{sec:Distributed-Implementation}}

In this section, we assume perfect PU channel sensing and infinitely
backlogged SUs, and focus on approaches based on policies in $\mathcal{C}_{0}$
that do not rely on central coordination in order to achieve the following
objective, OPT2: 
\begin{eqnarray}
\textrm{maximize} & \sum_{s\in\mathcal{S}}f_{s}\left(\bar{r}_{s}\right)\label{eq:dist_robj}\\
\mbox{subject to } & \text{ (\ref{eq:rate_prim_constr_3}), (\ref{eq:power_aver_constr_3}), (\ref{eq:sum_of_prob_3}), (\ref{eq:nonnegative1}) and (\ref{eq:aver_second1}) }\nonumber 
\end{eqnarray}
Functions $\left\{ f_{s}\left(\cdot\right)\right\} _{s\in\mathcal{S}}$
are usually selected so that certain fairness criteria for SU rate
allocation are satisfied, see \cite{Srikant2013} and \cite{A:Mo},
and they are assumed to be concave with respect to $\bar{r}_{s}$.
Thus, due to the fact that for all $s\in\mathcal{S}$, $\bar{r}_{s}$
is a linear function of variables $\left\{ \left\{ q\left(e,s,i\right)\right\} _{i\in\mathcal{I}_{s}^{0}},\left\{ q\left(b,s,i\right)\right\} _{i\in\mathcal{I}_{s}^{0}}\right\} $,
$f_{s}\left(\bar{r}_{s}\right)$ is also a concave function of these
variables. Hence, OPT2 is a convex optimization problem and can be
solved efficiently via interior point methods.

In an operational environment where parameters may change with time,
problem OPT2 will have to be solved whenever significant changes to
such parameters occur. A centralized solution requires a single node
to be responsible for gathering instantaneous parameter values, for
the solution of OPT2 and for determining the appropriate scheduling
of packet transmissions. While such a solution may be acceptable in
certain environments, it creates a ``single point of failure''.
Moreover the central node must be continually informing the SUs as
to which one will cooperate or transmit in each time slot and at which
power level. There may also be a scalability issue with this approach
since the number of variables is of the order $2\left|\mathcal{S}\right|\mathit{I}$,
where $I$ is the maximum number of power levels of SU nodes ($\sum_{i\in\mathcal{S}}\left|\mathcal{I}_{s}^{0}\right|$
parameters $\left\{ q\left(b,s,i\right)\right\} _{s\in\mathcal{S},\ i\in\mathcal{I}_{s}^{0}}$
plus $\sum_{i\in\mathcal{S}}\left|\mathcal{I}_{s}^{0}\right|$ parameters
$\left\{ q\left(e,s,i\right)\right\} _{s\in\mathcal{S},\ i\in\mathcal{I}_{s}^{0}}$).
Hence, depending on the computing power and memory availability at
the central node, solving problem OPT2 in a centralized location may
become prohibitive for larger number of SUs.

\subsubsection{Advantages of the Distributed Approach}

In this section, we derive a solution to OPT2 in a distributed fashion.
The main features of our approach are the following.

a) The PU involvement in the algorithm is only to announce its arrival
rate $\lambda_{p}$ at the beginning of the algorithm - no further
participation is required.

b) A SU node does not need to know the parameters (i.e., $r_{s}\left(i\right)$,
$r_{p}\left(s,i\right)$, $i\in\mathcal{I}_{s}$) of other SU nodes.

c) The distributed solution requires each SU node $s\in\mathcal{S}$
to solve optimization problems with $\left|\mathcal{I}_{s}^{0}\right|$
variables, hence the computational complexity per node does not increase
with the number of SU nodes.

d) Two messages are broadcasted by each SU node per iteration of the
distributed algorithm. The number of iterations for convergence depends
on the number of SU nodes, but this is tolerable for the algorithm
execution in a real-time setting.

e) Once convergence of the algorithm is reached for a given arrival
rate, the SUs need only observe the state of the PU channel (busy
or idle); they can decide autonomously which SU node is scheduled
to either cooperate with the PU, or to transmit its own traffic, without
the need of a scheduler, or the exchange of control messages.

We assume that there is a separate low-rate channel which is used
by the SUs for control message exchanges \cite{Lo}. In particular
we assume that control messages may be broadcasted among the SUs,
either because the low-rate channel is broadcast in nature, or through
the establishment of Broadcast Trees that usually are employed in
ad-hoc networks \cite{Ephrem,Papadimitriou}.

\subsubsection{Implementation of the Distributed Optimization Algorithm\label{sub:ADMoM-distributed}}

Towards a distributed solution to problem OPT2 we would ideally like
to decompose the global problem into $\left|\mathcal{S}\right|$ parallel
subproblems, each one involving only local variables and parameters
of node $s$. Among all alternatives we tried towards this end, the
best algorithm in terms of convergence was the one built upon the
\emph{Alternating Direction Method of Multipliers} (ADMoM), which
has superior convergence properties over the traditional dual ascent
method \cite{Boyd,D.P.Bertsekas_book,BertsekasTsitsiklis_book}. To
apply ADMoM to OPT2, we first turned the average power inequality
constraints (\ref{eq:power_aver_constr_3}) into equalities, by introducing
auxiliary variables $\left\{ y_{s}\right\} _{s\in\mathcal{S}}$, where
$y_{s}$ is associated with the respective $s^{th}$ constraint, and
is positive-valued. Also, for notational simplicity, we equivalently
rewrite problem OPT2 as OPT3 given by

\begin{eqnarray}
\textrm{minimize} & -\sum_{s\in\mathcal{S}}f_{s}\left(\phi_{s}\left(\boldsymbol{x}_{s}\right)\right)\label{eq:distr_obj_fin}\\
\textrm{subject to} & \sum_{s\in\mathcal{S}}g_{1s}\left(\boldsymbol{z}_{s}\right)=\lambda_{p}\label{eq:distr_rate_constr}\\
 & h_{s}\left(\boldsymbol{x}_{s},\boldsymbol{z}_{s},y_{s}\right)=\hat{P}_{s},\, s\in{\cal S}\label{eq:distr_power_constr}\\
 & \sum_{s\in\mathcal{S}}g_{2s}\left(\boldsymbol{x}_{s}\right)+\sum_{s\in\mathcal{S}}g_{2s}\left(\boldsymbol{z}_{s}\right)=1\label{eq:distr_prob_constr}\\
 & \boldsymbol{x}_{s}\geq0,\boldsymbol{z}_{s}\geq0,y_{s}\geq0,\, s\in{\cal S}\label{eq:distr_positivity_constr}
\end{eqnarray}
where we use the variables $\boldsymbol{x}_{s}\triangleq\left\{ q\left(e,s,i\right)\right\} _{i\in\mathcal{I}_{s}^{0}},\,\boldsymbol{z}_{s}\triangleq\left\{ q\left(b,s,i\right)\right\} _{i\in\mathcal{I}_{s}^{0}}$,
and we also define the following functions: $\phi_{s}\left(\boldsymbol{x}_{s}\right)\triangleq\sum_{i\in\mathcal{I}_{s}}r_{s}\left(i\right)q\left(e,s,i\right)$,
$g_{1s}\left(\boldsymbol{z}_{s}\right)\triangleq\sum_{i\in\mathcal{I}_{s}^{0}}r_{p}\left(s,i\right)q\left(b,s,i\right)$,
$g_{2s}\left(\boldsymbol{x}_{s}\right)\triangleq\mathbf{1}^{T}\boldsymbol{x}_{s}=\sum_{i\in{\cal I}_{s}^{0}}q\left(e,s,i\right)$,
$g_{2s}\left(\boldsymbol{z}_{s}\right)\triangleq\mathbf{1}^{T}\boldsymbol{z}_{s}=\sum_{i\in{\cal I}_{s}^{0}}q\left(b,s,i\right)$,
and 
\begin{equation}
h_{s}\left(\boldsymbol{x}_{s},\boldsymbol{z}_{s},y_{s}\right)\triangleq\sum_{i\in\mathcal{I}_{s}}P_{s}(i)q(e,s,i)+\sum_{i\in\mathcal{I}_{s}}P_{s}(i)q(b,s,i)+y_{s},\, s\in\mathcal{S}.
\end{equation}

Let $\nu$ and $\xi$ denote the dual variables associated with the
constraints of (\ref{eq:distr_rate_constr}) and (\ref{eq:distr_prob_constr})
respectively, and $\mu_{s}$ the dual variable associated with the
$s^{th}$ constraint of (\ref{eq:distr_power_constr}). Then, the
augmented Lagrange function corresponding to OPT3 used by ADMoM, parametrized
by the penalty parameter $\rho>0$, is given by \cite{Boyd,D.P.Bertsekas_book}
\begin{eqnarray}
L_{p} & = & \sum_{s\in\mathcal{S}}L_{s}-\nu\lambda_{p}-\xi+\frac{\rho}{2}\left\{ \left(\sum_{s\in\mathcal{\mathcal{S}}}g_{1s}\left(\boldsymbol{z}_{s}\right)-\lambda_{p}\right)^{2}\right.\label{eq:Aug_global_Lagrange}\\
 & + & \left.\sum_{s\in\mathcal{S}}\left(h_{s}\left(\boldsymbol{x}_{s},\boldsymbol{z}_{s},y_{s}\right)-\hat{P}_{s}\right)^{2}+\left(\sum_{s\in\mathcal{S}}g_{2s}\left(\boldsymbol{x}_{s}\right)+\sum_{s\in\mathcal{S}}g_{2s}\left(\boldsymbol{z}_{s}\right)-1\right)^{2}\right\} \nonumber 
\end{eqnarray}
with 
\begin{equation}
L_{s}\triangleq-f_{s}\left(\phi_{s}\left(\boldsymbol{x}_{s}\right)\right)+\nu g_{1s}\left(\boldsymbol{z}_{s}\right)+\mu_{s}\left(h_{s}\left(\boldsymbol{x}_{s},\boldsymbol{z}_{s},y_{s}\right)-\hat{P}_{s}\right)+\xi g_{2s}\left(\boldsymbol{x}_{s}\right)+\xi g_{2s}\left(\boldsymbol{z}_{s}\right),\, s\in{\cal S}.
\end{equation}

\emph{Computational complexity:} The optimization steps and variables
updates that need to be carried out at each SU node $s\in\mathcal{S}$,
according to ADMoM, are given by 
\begin{eqnarray}
\boldsymbol{x}_{s}^{k+1} & = & \arg\min_{\boldsymbol{x}_{s}\geq0}L_{s}\left(\boldsymbol{x}_{s},\boldsymbol{z}_{s}^{k},y_{s}^{k},v^{k},\xi^{k},\mu_{s}^{k}\right)+\frac{\rho}{2}\left(h_{s}\left(\boldsymbol{x}_{s},\boldsymbol{z}_{s}^{k},y_{s}^{k}\right)-\hat{P_{s}}\right)^{2}\label{eq:update_x}\\
 & + & \frac{\rho}{2}\left(\sum_{m=1}^{s-1}g_{2m}\left(\boldsymbol{x}_{m}^{k+1}\right)+\sum_{m=s+1}^{\mid\mathcal{S\mid}}g_{2m}\left(\boldsymbol{x}_{m}^{k}\right)+g_{2s}\left(\boldsymbol{x}_{s}\right)+\sum_{s\in\mathcal{S}}g_{2s}\left(\boldsymbol{z}_{s}^{k}\right)-1\right)^{2},\nonumber 
\end{eqnarray}
\begin{eqnarray}
\boldsymbol{z}_{s}^{k+1} & = & \arg\min_{\boldsymbol{z}_{s}\geq0}L_{s}\left(\boldsymbol{x}_{s}^{k+1},\boldsymbol{z}_{s},y_{s}^{k},v^{k},\xi^{k},\mu_{s}^{k}\right)+\frac{\rho}{2}\left(h_{s}\left(\boldsymbol{x}_{s}^{k+1},\boldsymbol{z}_{s},y_{s}^{k}\right)-\hat{P_{s}}\right)^{2}\label{eq:update_z}\\
 & + & \frac{\rho}{2}\left(\sum_{m=1}^{s-1}g_{1m}\left(\boldsymbol{z}_{m}^{k+1}\right)+\sum_{m=s+1}^{\mid\mathcal{S\mid}}g_{1m}\left(\boldsymbol{z}_{m}^{k}\right)+g_{1s}\left(\boldsymbol{z}_{s}\right)-\lambda_{p}\right)^{2}\nonumber \\
 & + & \frac{\rho}{2}\left(\sum_{s\in\mathcal{S}}g_{2s}\left(\boldsymbol{x}_{s}^{k+1}\right)+\sum_{m=1}^{s-1}g_{2m}\left(\boldsymbol{z}_{m}^{k+1}\right)+\sum_{m=s+1}^{\mid\mathcal{S\mid}}g_{2m}\left(\boldsymbol{z}_{m}^{k}\right)+g_{2s}\left(\boldsymbol{z}_{s}\right)-1\right)^{2},\nonumber 
\end{eqnarray}
\begin{equation}
y_{s}^{k+1}=\arg\min_{y_{s}\geq0}L_{s}\left(\boldsymbol{x}_{s}^{k+1},\boldsymbol{z}_{s}^{k+1},y_{s},v^{k},\xi^{k},\mu_{s}^{k}\right)+\frac{\rho}{2}\left(h_{s}\left(\boldsymbol{x}_{s}^{k+1},\boldsymbol{z}_{s}^{k+1},y_{s}\right)-\hat{P_{s}}\right)^{2},\label{eq:update_y}
\end{equation}
\begin{equation}
\xi^{k+1}=\xi^{k}+\rho\left(\sum_{s\in\mathcal{S}}g_{2s}\left(\boldsymbol{x}_{s}^{k+1}\right)+\sum_{s\in\mathcal{S}}g_{2s}\left(\boldsymbol{z}_{s}^{k+1}\right)-1\right),\label{eq:update_ji-1}
\end{equation}
\begin{equation}
\nu^{k+1}=\nu^{k}+\rho\left(\sum_{s\in\mathcal{S}}g_{1s}\left(\boldsymbol{z}_{s}^{k+1}\right)-\lambda_{p}\right),\label{eq:update_v}
\end{equation}
\begin{equation}
\mu_{s}^{k+1}=\mu_{s}^{k}+\rho\left(h_{s}\left(\boldsymbol{x}_{s}^{k+1},\boldsymbol{z}_{s}^{k+1},y_{s}^{k+1}\right)-\hat{P_{s}}\right),\label{eq:update_mi}
\end{equation}
where $k$ denotes the iteration index. Note that the computational
burden is distributed across SU nodes; the computational complexity
at each node depends primarily on the two quadratic optimization problems
in (\ref{eq:update_x}) and (\ref{eq:update_z}), each of which has
$\left|\mathcal{I}_{s}^{0}\right|$ variables, and can be efficiently
solved via interior point methods, or standard methods such as Newton
Method. All the following steps involve a single variable and are
straightforward.

\emph{Communication overhead:} Each node $s$, in order to perform
the steps in (\ref{eq:update_x}) and (\ref{eq:update_z}), needs
to know information concerning the updated local variables of other
nodes. This can be accomplished through message broadcasts by each
SU node via the control channel in the following manner. The nodes
update their local variables and broadcast the messages required sequentially,
in a prespecified order. Specifically, for the step in (\ref{eq:update_x}),
each node $s\in\mathcal{S}$ updates its primal variable $\boldsymbol{x}_{s}^{k+1}$
and broadcasts message $g_{2s}\left(\boldsymbol{x}_{s}^{k+1}\right)$.
Similarly, for the step in (\ref{eq:update_z}), each SU node updates
its variable $\boldsymbol{z}_{s}^{k+1}$and broadcasts $g_{1s}\left(\boldsymbol{z}_{s}^{k+1}\right)$
and $g_{2s}\left(\boldsymbol{z}_{s}^{k+1}\right)$ in one message,
according to the prespecified order. Steps dictated by (\ref{eq:update_y})-(\ref{eq:update_mi}),
for each node $s$, require only its local variables and information
that is already acquired by $s$ from the previous message broadcasts
and thus can be implemented in parallel by all nodes. Each iteration
of the distributed algorithm consists of one round of these update
steps by all $\left|\mathcal{S}\right|$ nodes. Consequently, the
communication overhead of the algorithm is $2\left|\mathcal{S}\right|$
message broadcasts per iteration.

\emph{Convergence:} For the convergence of the algorithm in decentralized
manner, each SU keeps track of a local metric and determines local
convergence with respect to it, within a prespecified accuracy. This
local metric for each node $s\in\mathcal{S}$ may be the the successive
differences of its local objective function under optimization, i.e.,
$f_{s}\left(\boldsymbol{x}_{s}\right)$. Once this local metric drops
under the prespecified accuracy, local convergence is declared, and
node $s$ announces it via the control channel. As soon as all SU
nodes reach convergence, the algorithm terminates.

\emph{Real-time implementation: }We assume that the PU broadcasts
its average arrival rate $\lambda_{p}$ at the beginning of the algorithm.
Once convergence of the algorithm for a given $\lambda_{p}$ is reached,
all SUs have knowledge of the sums of probabilities $g_{2s}\left(\boldsymbol{x}_{s}^{opt}\right),g_{2s}\left(\boldsymbol{z}_{s}^{opt}\right),\forall s\in\mathcal{S}$.
Thus, if the SUs use the same randomization algorithm and common seed,
as long as they observe the state of the PU channel, they can all
independently produce the same result as to who SU is scheduled to
cooperate with the PU or transmit its own data in every time slot.
Then, the scheduled SU determines its power level for its transmission
based on its own probability parameters. The system evolves without
the need for further coordination among network nodes.

The algorithm runs again only when some of the parameters of the operational
environment change significantly. Thus, when the arrival rate changes
within a pre-specified percentage of its previous value, the PU informs
the SUs about the new value of $\lambda_{p}$. Also, in case wireless
channel gains change for some SU within a certain percentage, the
corresponding SU may announce the rerun of the algorithm. In such
cases the algorithm can adapt to changes in the operational environment;
the problem is not solved from scratch, but the algorithm is initialized
at the optimal point of the previous system state. This speeds up
its convergence and reduces the overall communication overhead, as
will be shown in the simulation results that follow.

\emph{Exogenous Packet arrivals to SU queues:} In case of this scenario,
we seek a decentralized solution to the optimization problem (\ref{eq:throughputOptimize})
according to subsection \ref{sec:Incorporating-Exogenous-Packet}.
However, if $f(\boldsymbol{R})$ is separable, i.e., $f(\boldsymbol{R})=\sum_{s\in{\cal S}}f_{s}\left(R_{S}\right)$,
then problem in (\ref{eq:throughputOptimize}) is essentially identical
to the one in (\ref{eq:dist_robj}) where we replace $f_{s}(\bar{r}_{s})$
with $f_{s}\left(\min\{\lambda_{s},\bar{r}_{s}\}\right)$. We can
therefore employ ADMoM using the same techniques as previously to
provide a distributed implementation of the current optimization problem.
Note that the fact that in the distributed implementation only SU
$s$ needs to know $f_{s}\left(\min\left\{ \lambda_{s},r_{s}\right\} \right)$,
implies that each SU needs to know only its arrival rate in order
to implement the distributed algorithm.

\section{\label{sec:Simulation-Results}Simulation and Numerical Results}

In this section, we confirm the optimality claims in terms of performance
for the proposed class of policies through several simulation experiments
for different scenarios. First, we assume that SUs are infinitely
backlogged and spectrum sensing is perfect. In this scenario, the
performance of an optimal policy in ${\cal C}_{0}$ is compared to
the transmission algorithm presented in \cite{Neely_cooper} and an
optimal dynamic policy from ${\cal C}_{2}$, constructed through the
Lyapunov optimization techniques \cite{B:Neely_book}. Furthermore,
the convergence of the distributed algorithm, as well as its ability
to adapt to changing parameters is studied. Next, we consider exogenous
packet arrivals to SUs queues and the performance of an optimal policy
in the proposed class $\tilde{{\cal C}}_{0}$ is presented. Finally,
imperfect spectrum sensing is assumed and the convexity of the resulting
optimization problem is investigated. In all the above scenarios,
we consider a system model with one PU and several SUs, and as objective
optimization function $f\left(\cdot\right)$ the sum of transmission
rates of the SUs, i.e., $f\left(\bar{\boldsymbol{r}}\right)=\sum_{s\in\mathbb{S}}\bar{r}_{s}$.

\begin{figure}
\centering\includegraphics[width=0.54\columnwidth]{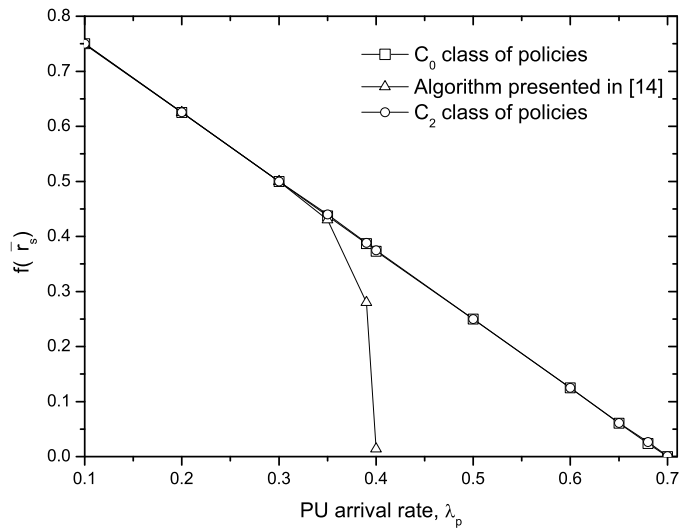}
\protect\protect\caption{The SU throughput utility function.}

\label{Fig:rate_2_scenario} 
\end{figure}

\begin{figure}
\centering\includegraphics[width=0.54\columnwidth]{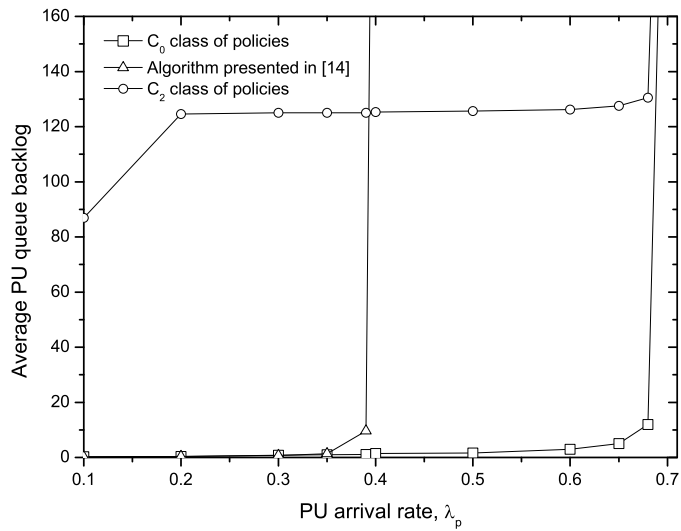}
\protect\protect\caption{The average backlog of the PU queue.}

\label{Fig:delay_2_scenario} 
\end{figure}

Assuming perfect sensing and infinitely backlogged SUs, the performance
of a setup which consists of $5$ SUs and a set of $5$ available
transmit power levels is investigated in Figs. \ref{Fig:rate_2_scenario}-\ref{Fig:delay_2_scenario},
in terms of $f\left(\mathbf{\bar{r}}\right)$ and average backlog
of PU queue. Specifically, we assume for this setup that $\mathcal{I}_{s}^{0}=\left\{ 0,1,2,3,4\right\} $,
$\mathbf{P}_{s}=\left\{ 0,0.25,0.5,0.75,1\right\} $, $r_{p}\left(0\right)=0.4$,
$r_{p}\left(s,1\right)=0.5$, $r_{p}\left(s,2\right)=0.6$, $r_{p}\left(s,3\right)=0.7$,
$r_{p}\left(s,4\right)=0.8$, $r_{s}\left(1\right)=0.3$, $r_{s}\left(2\right)=0.5$,
$r_{s}\left(3\right)=0.8$, $r_{s}\left(4\right)=1$, and the average
power constraint is $\hat{P}_{s}=0.15$, for all $s\in\mathcal{S}$.
It can be seen in Fig. \ref{Fig:rate_2_scenario} that the sum rate
achieved by SUs that employ an optimal policy from the restricted
class of policies ${\cal C}_{0}$ is identical to the sum rate achieved
under the optimal policy in ${\cal C}_{2}$. This is in accordance
with the main result of Theorem \ref{thm:Equivalence}. Additionally,
as it is illustrated by Fig. \ref{Fig:delay_2_scenario}, the average
backlog of the PU queue remains very low under the optimal policy
in $\mathcal{C}_{0}$.

On the contrary, the dynamic policy from $\mathcal{C}_{2}$ induces
large sizes to PU queue even for small arrival rates. Moreover, when
compared to the control algorithm presented in \cite{Neely_cooper},
the class ${\cal C}_{0}$ of policies extends the range of $\lambda_{p}$
that can be supported by the system, \emph{providing mutual benefits
to both PU and SUs out of their cooperation}. In particular, transmission
rates higher than the PU queue service rate without SU cooperation
can be supported for the PU through the class of policies $\mathcal{C}_{0}$,
while transmission opportunities are provided to SUs to transmit their
own data. It should be noted that the policy in \cite{Neely_cooper}
was shown to be optimal for $\lambda_{p}<0.4$, and this is confirmed
in Fig. \ref{Fig:rate_2_scenario}, where it is shown that all three
policies achieve the same sum-rate for $\lambda_{p}<0.4$. However,
the policy in \cite{Neely_cooper} renders the PU queue unstable for
$\lambda_{p}>0.4$ and reduces the SU sum rates to zero. The reason
is the following. In \cite{Neely_cooper}, decisions are taken at
the end of busy periods of the PU queue. If $\lambda_{p}>0.4$, whenever
a decision not to cooperate is taken, there is a nonzero probability
that the primary queue never becomes empty, and hence there is no
possibility for the SUs to take corrective actions.

For the same scenario and system setup, we also evaluate the performance
of the proposed distributed algorithm. Regarding the distributed implementation
parameters, we set the desired accuracy for convergence equal to $\epsilon=10^{-5}$,
while the penalty parameter is taken to be $\rho=0.1$. For the arbitrary
initialization of the algorithm, we used $\left\{ q\left(e,s,i\right)^{0}\right\} _{i\in\mathcal{I}_{s}^{0}}=0.01,\forall s\in\mathcal{S}$,
$\left\{ q\left(b,s,i\right)^{0}\right\} _{i\in\mathcal{I}_{s}^{0}}=0.03,\forall s\in\mathcal{S}$,
$\left\{ \mu_{s}^{0}\right\} _{s\in\mathcal{S}}=1$, $\xi^{0}=1$,
$\nu^{0}=1$. The distributed algorithm was tested against the centralized
solution to problem OPT2, in terms of the value of the objective,
and for various values of the PU arrival rate $\lambda_{p}$. It was
observed that the numerical results obtained from both centralized
and distributed implementations were identical (equal with those provided
by Fig. \ref{Fig:rate_2_scenario}); this shows that our proposed
algorithm keeps up with its centralized counterpart, which can be
justified by the convergence properties of ADMoM. Regarding the convergence
speed, the number of iterations required for convergence within the
given accuracy are given in Table \ref{Tab_2}, when the arrival rate
$\lambda_{p}$ is varied inside the stability region and the proposed
algorithm begins from scratch (arbitrary initialization). Obviously
the algorithm is efficient enough, since it converges within a tolerable
number of iterations for low PU transmission rates, while the convergence
is even faster at higher ones. This can be explained by the fact that
as $\lambda_{p}$ increases, the constraints in (\ref{eq:distr_rate_constr})-(\ref{eq:distr_positivity_constr})
get tighter, restricting the feasibility set of the problem variables
$\left\{ \boldsymbol{x}_{s},\boldsymbol{z}_{s},y_{s}\right\} _{s\in\mathcal{S}}$.
Consequently, since the distributed algorithm searches for the optimal
solution within the feasibility set in each case of $\lambda_{p}$,
it needs more iterations to converge when searching within a wider
set than when searching within a narrower set. Finally, the adaptivity
of the distributed algorithm to changes in the arrival rate $\lambda_{p}$,
is investigated in Table \ref{Tab3}. In particular, we begin with
an initial rate equal to $\lambda_{p}^{0}=0.5$, and run the algorithm
from scratch, as described above. For all values of $\lambda_{p}$
different from $\lambda_{p}^{0}$, we use as initialization for the
algorithm the optimal point found at $\lambda_{p}^{0}$, and write
down the number of iterations required for convergence within the
given accuracy. Clearly, there is a remarkable reduction in the total
number of iterations required for convergence compared with the arbitrary
initialization.

\begin{table}[t]
\protect\protect\caption{Number of iterations for the distributed algorithm.}

\centering%
\begin{tabular}{|c|c|c|c|c|c|c|}
\hline 
$\lambda_{p}$  & 0.2  & 0.3  & 0.4  & 0.5  & 0.6  & 0.7 \tabularnewline
\hline 
\hline 
$\sharp$ of iterations  & 263  & 172  & 129  & 119  & 105  & 74 \tabularnewline
\hline 
\end{tabular}\label{Tab_2} 
\end{table}

\begin{table}[t]
\protect\protect\caption{Number of iterations as PU rate changes from $\lambda_{p}^{0}=0.5$
to $\lambda_{p}$.}

\centering%
\begin{tabular}{|c|c|c|c|c|c|c|c|}
\hline 
$\lambda_{p}$  & 0.35  & 0.4  & 0.45  & 0.52  & 0.55  & 0.6  & 0.7\tabularnewline
\hline 
\hline 
$\sharp$ of iterations  & 44  & 34  & 39  & 29  & 39  & 45  & 16\tabularnewline
\hline 
\end{tabular}\label{Tab3} 
\end{table}

Next, we additionally consider exogenous SU packet arrivals to the
to the system setup described above. For this scenario, the throughput
performance of the optimal policies in class $\tilde{{\cal C}}_{0}$
is investigated for the cases where $\sum_{s\in\mathcal{S}}\lambda_{s}$
is either well within or outside the achievable rate region $\mathcal{R}_{0}$,
for both centralized and distributed implementations. Specifically,
we initially fix $\sum_{s\in\mathcal{S}}\lambda_{s}$ inside the achievable
rate region for each case of $\lambda_{p}$ considered; $\lambda_{p}$
varies in the range $\left[0.2,\ldots,0.7\right]$, while $\sum_{s\in\mathcal{S}}\lambda_{s}$
is fixed equal to $0.05$, where $\lambda_{s}=0.01$, for all $s\in\mathcal{S}$.
It was observed that the optimization objective values attained from
both implementations are identical and equal to $\sum_{s\in\mathcal{S}}\lambda_{s}$,
for each value of $\lambda_{p}$. Secondly, we consider $\sum_{s\in\mathcal{S}}\lambda_{s}$
outside the achievable rate region $\mathcal{R}_{0}$ for each value
of the PU arrival rate $\lambda_{p}$; $\lambda_{p}$ varies in the
range $\left[0.2,\ldots,0.7\right]$, while $\sum_{s\in\mathcal{S}}\lambda_{s}$
is fixed and equal to $1$, where $\lambda_{s}=0.2$, for all $s\in\mathcal{S}$.
It was observed that the respective throughput utility that results
from both centralized and distributed implementations coincide and
are equal with the corresponding results when the SU queues are infinitely
backlogged (provided by Fig. \ref{Fig:rate_2_scenario}). Hence, the
optimal policies in class $\tilde{{\cal C}}_{0}$ achieve the maximum
possible value for the SU throughput utility function. The number
of iterations required for the convergence of the distributed algorithm
is shown in Tables \ref{Tab4} and \ref{Tab5}. For the derivation
of these results, an accuracy of $\epsilon=10^{-5}$ is assumed and
the distributed algorithm runs from scratch for each value of $\lambda_{p}$
considered, while using the same initialization values for its variables
as those used in the simulation experiments concerning the first scenario.
The distributed algorithm converges again within a tolerable number
of iterations.

\begin{table}
\protect\protect\caption{Number of iterations for the distributed algorithm, $\sum_{s\in\mathcal{S}}\lambda_{s}=0.05$.}

\centering%
\begin{tabular}{|c|c|c|c|c|c|c|}
\hline 
$\lambda_{p}$  & 0.2  & 0.3  & 0.4  & 0.5  & 0.6  & 0.7\tabularnewline
\hline 
\hline 
$\sharp$ of iterations  & 93  & 89  & 95  & 137  & 301  & 227\tabularnewline
\hline 
\end{tabular}\label{Tab4} 
\end{table}

\begin{table}
\protect\protect\caption{Number of iterations for the distributed algorithm, $\sum_{s\in\mathcal{S}}\lambda_{s}=1$.}

\centering%
\begin{tabular}{|c|c|c|c|c|c|c|}
\hline 
$\lambda_{p}$  & 0.2  & 0.3  & 0.4  & 0.5  & 0.6  & 0.7\tabularnewline
\hline 
\hline 
$\sharp$ of iterations  & 268  & 127  & 136  & 116  & 103  & 72\tabularnewline
\hline 
\end{tabular}\label{Tab5} 
\end{table}

\begin{figure}
\centering\includegraphics[width=0.54\columnwidth]{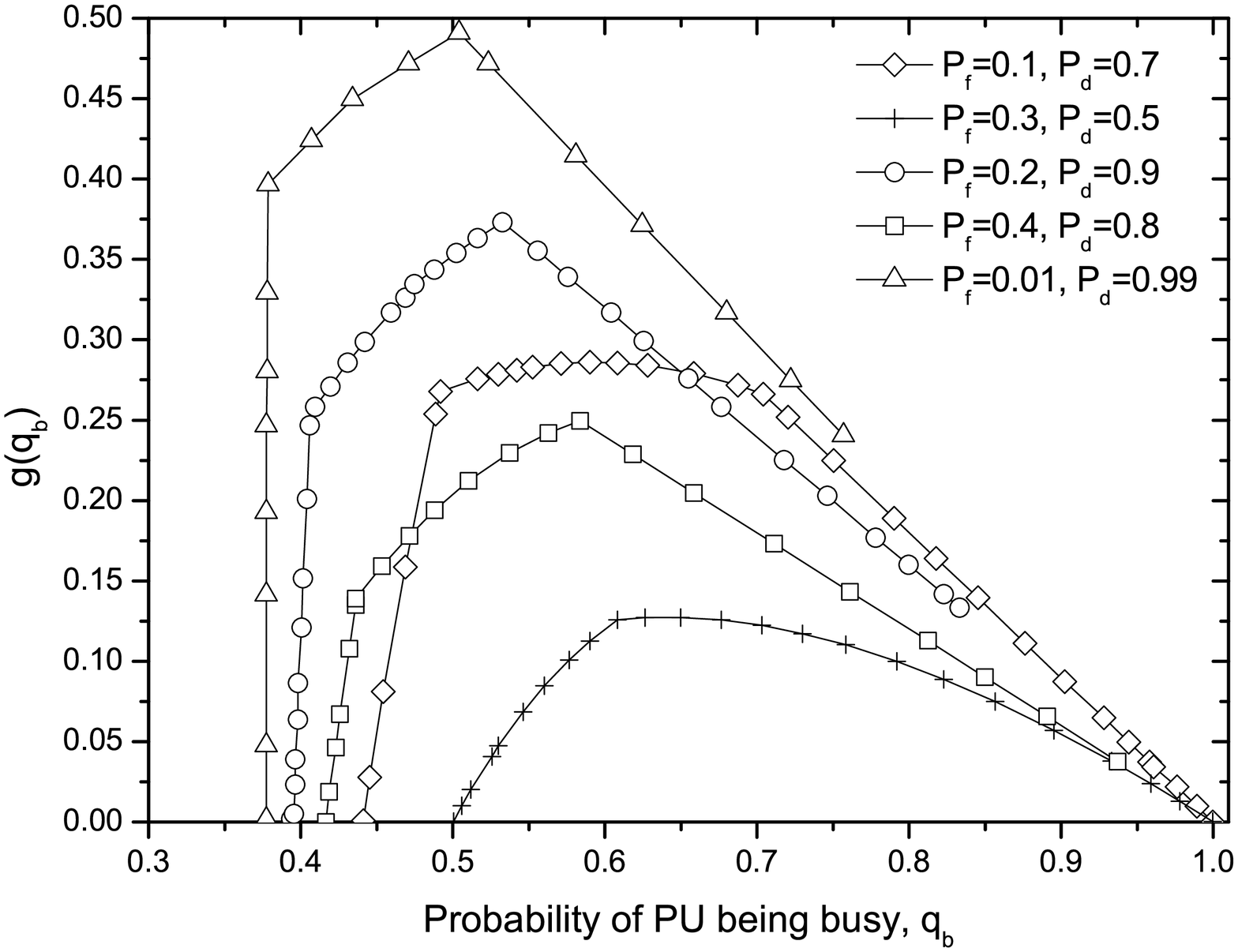} \protect\protect\caption{Imperfect sensing effects.}

\label{Fig:im1} 
\end{figure}

Finally, the effects of imperfect spectrum sensing are investigated
in Fig. \ref{Fig:im1}. Specifically, assuming the same system setup
and $\lambda_{p}=0.3$, we solve numerically OPT1, by fixing $q_{b}$
and calculating the maximum value of the objective of OPT1 $g(q_{b})$
when $q_{b}\in\left[0,1\right]$, for various values of $\mathcal{P}_{D}$
and $\mathcal{P}_{F}$. It can be observed that $q_{b}$ takes values
only on the interval specified by the proposition \ref{Rem:Regio},
for all values of $\mathcal{P}_{D}$ and $\mathcal{P}_{F}$ considered;
thus, restricting the region of $q_{b}$ where exhaustive linear search
methods have to search. Furthermore, when investigating the concavity
of $g(q_{b})$, simulation results indicate that $g(q_{b})$ is concave
with respect to $q_{b}$, irrespective of the values of $\mathcal{P}_{D}$
and $\mathcal{P}_{F}$ considered. As discussed in section \ref{sec:Imperfect-Sensing},
if this property is true in general, then the computational complexity
of the centralized solution, as well as the computational complexity
and overhead of a potential distributed implementation, can be significantly
reduced.

\section{\label{sec:Conclusions}Conclusions}

In this work we propose and investigate novel primary-secondary user
cooperation policies for cognitive radio networks that orchestrate
a PU and co-existing SUs in a wireless channel. The key idea is that
SUs increase the service rate of the PU queue and therefore they increase
the range of arrival rate of the PU for which its queue is stable.
At the same time, the PU queue empties more often, and therefore the
channel becomes idle more often, thus giving to SUs more transmission
opportunities. Our major contribution to the state of the art is the
proposition of policies that require only the state of PU channel
(busy or empty) for their implementation, yet: 1) they achieve substantial
augmentation of the stability region of the PU queue, and 2) they
can obtain any long term SU rates achievable by policies for which
the restriction of always giving priority to PU traffic is removed.
The mode of operation, the performance space and the optimality of
the proposed policies is investigated in models where SUs are either
infinitely backlogged, or finite exogenous packet arrivals to SU queues
occur. An important feature of the proposed transmission algorithm
is that the optimal transmit probabilities can be computed offline,
through solving a convex optimization problem, and can be communicated
to users. A centralized and a distributed version of the algorithm
are presented, both of which are applicable depending on the setup.
Simulation results verify the benefits of our approach, as well as
the consistency of the proposed distributed algorithm with its centralized
counterpart performance-wise. A possible extension to this work is
the design of a dynamic, online version of the proposed algorithm.
Furthermore, the uncoordinated interaction of multiple PUs and SUs
gives rise to game-theoretic models that warrant further investigation.

\appendices{}

\section{Proof of Proposition \ref{thm:Equivalence}\label{sec:Proof-of-Equivalence}}

Let us define as ${\cal R}_{0}$ the performance space of $\bar{r}_{s}$
defined by (\ref{eq:h1}) where $q_{b},$ $q_{e},$ $\left\{ q\left(s,i\left|e\right.\right)\right\} $,
$\left\{ q\left(s,i\left|b\right.\right)\right\} $ satisfy (\ref{eq:rate_equality})-(\ref{eq:non_negativity})
and ${\cal \hat{R}}_{0}$ the performance space of $\bar{r}_{s}$
defined by (\ref{eq:aver_second1}) where $\left\{ q\left(e,s,i\right)\right\} $,
$\left\{ q\left(b,s,i\right)\right\} $ satisfy (\ref{eq:rate_prim_constr_3})-(\ref{eq:nonnegative1}).
Due to the transformation, it holds that any $\bar{r}_{s}\in\mathcal{R}_{0}$
is also in ${\cal \hat{R}}_{0},$ i.e., $\mathcal{R}_{0}\subseteq{\cal \hat{R}}_{0}$.

Conversely, we consider any $\bar{r}_{s}\in{\cal \hat{R}}_{0}.$ Assuming
that $q_{e}\neq0$ and $q_{b}\neq0$, we make the transformation $q_{e}=\sum_{s\in\mathcal{S}}\sum_{i\in\mathcal{I}_{s}^{0}}q\left(e,s,i\right),$
$q_{b}=\sum_{s\in\mathcal{S}}\sum_{i\in\mathcal{I}_{s}^{0}}q\left(b,s,i\right),$
$q\left(s,i\left|e\right.\right)=\frac{q\left(e,s,i\right)}{q_{e}}$
and $q(s,i\left|b\right.)=\frac{q(b,s,i)}{q_{b}}.$ Since the parameters
$\left\{ q\left(e,s,i\right)\right\} $ and $\left\{ q\left(b,s,i\right)\right\} $
satisfy (\ref{eq:rate_prim_constr_3})-(\ref{eq:nonnegative1}), it
can be shown after some basic algebraic manipulations that $q_{b},$
$q_{e},$ $\left\{ q\left(s,i\left|e\right.\right)\right\} $ and
$\left\{ q(s,i\left|b\right.)\right\} $ satisfy (\ref{eq:rate_equality})-(\ref{eq:non_negativity}).
Hence, $\bar{r}_{s}\in\mathcal{R}_{0},$ i.e., ${\cal \hat{R}}_{0}\subseteq\mathcal{R}_{0}$.

In case that $q_{b}=0$, we define $q(s,i\left|b\right.)=0$ for $s\in{\cal S}$
and $i\in{\cal I}_{s}^{0}$. Again after some basic algebraic manipulations,
it can be shown that $\mathcal{\hat{R}}_{0}\subseteq\mathcal{R}_{0}$.
Similarly, when $q_{e}=0$, we define $q(s,i\left|e\right.)=0$ for
$s\in{\cal S}$ and $i\in{\cal I}_{s}^{0}$ and it can be shown that
${\cal \hat{R}}_{0}\subseteq\mathcal{R}_{0}$.

Based on the above, it can be concluded that $\mathcal{R}_{0}={\cal \hat{R}}_{0}$.

\section{Proof of Corrolary \ref{cor:The-stability-region-1}\label{sec:Proof-of-max_rate}}

The optimization problem defined in the corollary has always a feasible
solution, which can be obtained through setting $x\left(b,s,i\right)=0$
for $s\in\mathcal{S}$, $i\in\mathcal{I}_{s}$ and selecting arbitrarily
$x(b,s,0)\geq0$, so that $\sum_{s}x(b,s,0)=1$, resulting to $\sum_{s\in\mathcal{S}}\sum_{i\in\mathcal{I}_{s}^{0}}r_{p}(s,i)x(b,s,i)=r_{p}(0).$
Since $\hat{\lambda}$ is the optimal value of its objective, it follows
that $r_{p}\left(0\right)\leq\hat{\lambda}$ as expected. Physically,
this choice of parameters, corresponds to the case where SUs never
cooperate.

If $\lambda_{p}$ belongs to the stability region of the system, then
(\ref{eq:rate_prim_constr_3})-(\ref{eq:nonnegative1}) are satisfied.
But then, Eqs. (\ref{eq:max_rate_equality-1})-(\ref{eq:nonnegative3-1})
are also satisfied by choosing $x(b,s,i)=q(b,s,i)$, which implies
that $\lambda_{p}\leq\hat{\lambda}$.

Conversely, given any $\lambda_{p}\leq\hat{\lambda}$, the choice
of $q(b,s,i)=\left(\lambda_{p}/\hat{\lambda}\right)\hat{x}\left(b,s,i\right)$
for $s\in\mathcal{S}$ and $i\in\mathcal{I}_{s}^{0}$, $q(e,s,i)=0$
for $s\in\mathcal{S}$ and $i\in\mathcal{I}_{s}$, and $q(e,s,0)\geq0$
arbitrarily chosen so that $\sum_{s\in\mathcal{S}}q(e,s,0)=1-\sum_{s\in\mathcal{S}}\sum_{i\in\mathcal{I}_{s}^{0}}q(b,s,i)$
satisfies (\ref{eq:power_aver_constr_3})-(\ref{eq:nonnegative1}).
In addition, $\sum_{s\in\mathcal{S}}\sum_{i\in\mathcal{I}_{s}^{0}}r_{p}(s,i)q(b,s,i)=\lambda_{p},$
proving that the $\lambda_{p}$ belongs to the stability region of
the PU queue. This concludes the proof.

\section{Proof of Theorem \ref{thm:Equivalence2} \label{sec:Proof-of-Equivalence2}}

Let $\mathbf{\bar{r}}\in\mathcal{R}_{2}$. If $\lambda_{p}=\sum_{s\in\mathcal{S}}\sum_{i\in\mathcal{I}_{s}^{0}}r_{p}(s,i)p(1,s,i),$
then clearly $\mathbf{\bar{r}}\in\mathcal{R}_{0}$. Assume next that
$\lambda_{p}<\sum_{s\in\mathcal{S}}\sum_{i\in\mathcal{I}_{s}^{0}}r_{p}(s,i)p(1,s,i)$.
We distinguish the following cases:

\emph{Case 1}. $\lambda_{p}\geq r_{p}\left(0\right)p(1),$ where $p\left(1\right)\triangleq\sum_{s\in\mathcal{S}}\sum_{i\in\mathcal{I}_{s}^{0}}p(1,s,i)$
denotes the total probability that PU transmits, summed over all SU
$s$ and transmit power levels.

Note that since $r_{p}\left(0\right)p\left(1\right)\leq\lambda_{p}<\sum_{s\in\mathcal{S}}\sum_{i\in\mathcal{I}_{s}^{0}}r_{p}(s,i)p(1,s,i),$
for each $\lambda_{p}$ in the interval above, there exists a parameter
$\alpha$, with $0\leq\alpha<1$, such that it holds 
\begin{equation}
\lambda_{p}=\alpha\left(\sum_{s\in\mathcal{S}}\sum_{i\in\mathcal{I}_{s}}r_{p}(s,i)p(1,s,i)\right)+\left(1-\alpha\right)r_{p}\left(0\right)p\left(1\right).\label{eq:Lequality}
\end{equation}
We define now the new set of parameters $q\left(b,s,i\right)$ and
$q\left(e,s,i)\right)$ by setting $q\left(e,s,i\right)=p\left(0,s,i\right)$
for all $s\in\mathcal{S}$ and $i\in\mathcal{I}_{s}^{0}$ and 
\begin{equation}
q(b,s,i)=\left\{ \begin{array}{cc}
\alpha p(1,s,i) & \mbox{if }i\in\mathcal{I}_{s}\\
\alpha p(1,s,0)+(1-\alpha)p\left(1,s\right) & \mbox{if }i=0,
\end{array}\right.
\end{equation}
for all $s\in{\cal S}$, where $p\left(1,s\right)\triangleq\sum\limits _{j\in{\cal I}_{s}^{0}}p(1,s,j)$.
Since $0\leq\alpha<1$, parameters $q\left(e,s,i\right)$ and $q\left(b,s,i\right)$,
for all $s\in\mathcal{S}$ and $i\in\mathcal{I}_{s}^{0}$, are non-negative.
Furthermore, note that $\sum_{i\in\mathcal{I}_{s}^{0}}q(b,s,i)=\sum_{i\in\mathcal{I}_{s}^{0}}p(1,s,i)$.
Hence the new set of parameters satisfies (\ref{eq:sum_of_prob_3}).
Also, since $P_{s}\left(0\right)=0$, it can be shown that the new
set of parameters satisfy (\ref{eq:aver_power-2}). Finally, due to
(\ref{eq:Lequality}), it follows that (\ref{eq:rate_prim_constr_3})
is satisfied. Hence the new set of parameters satisfy (\ref{eq:rate_prim_constr_3})-(\ref{eq:nonnegative1}).
Also since the SU rates computed according to (\ref{eq:aver_second1})
(where $q\left(e,s,i\right)=p\left(0,s,i\right)$ for all $s\in\mathcal{S}$
and $i\in\mathcal{I}_{s}^{0}$) are the same as the ones given by
(\ref{eq:SecondaryRate}), it follows that $\mathbf{\bar{r}}\in\mathcal{R}_{0}$.

\emph{Case 2. }$\lambda_{p}<r_{p}\left(0\right)p(1)$. Define the
new set of parameters as follows 
\begin{equation}
q\left(b,s,i\right)=\left\{ \begin{array}{cc}
0 & \mbox{if \ensuremath{i\in\mathcal{I}_{s}}}\\
\frac{\lambda_{p}}{r_{p\left(0\right)}p\left(1\right)}p\left(1,s\right) & i=0,
\end{array}\right.
\end{equation}
and 
\begin{equation}
q\left(e,s,i\right)=\left\{ \begin{array}{cc}
p\left(0,s,i\right) & \mbox{if \ensuremath{i\in\mathcal{I}_{s}}}\\
\beta\sum\limits _{i\in\mathcal{I}_{s}^{0}}p\left(0,s,i\right)+p\left(0,s,0\right) & \mbox{if \ensuremath{i=0},}
\end{array}\right.\label{eq:newq_esi}
\end{equation}
for all $s\in{\cal S}$, where $\beta=\frac{1-\frac{\lambda_{p}}{r_{p}\left(0\right)}}{1-p(1)}-1$.
Since $\lambda_{p}<r_{p}\left(0\right)p(1)$, and $p\left(1\right)\leq1$,
it follows that $\beta>0,$ hence, all the defined parameters are
non-negative. Also, due to (\ref{eq:sum_prob_busy-2}), (\ref{eq:sum_of_prob_3})
is satisfied. Next, it can be easily shown that (\ref{eq:rate_prim_constr_3})
is satisfied. Furthermore, due to (\ref{eq:aver_power-2}), (\ref{eq:power_aver_constr_3})
is also satisfied. Finally, since $P_{s}\left(0\right)=0,$ it follows
that the SU rates computed according to (\ref{eq:aver_second1}) and
(\ref{eq:newq_esi}), are the same as the ones given by (\ref{eq:SecondaryRate}).
Hence we conclude that $\mathbf{\bar{r}}\in\mathcal{R}_{0}$.

\section{Proof of Proposition \ref{Rem:Regio}\label{sec:Proof-of-Remark}}

We assume first that there exist $q_{b}$, $\left\{ q\left(s,i\left|b\right.\right)\right\} $
and $\left\{ q\left(s,i\left|e\right.\right)\right\} $ that satisfy
the constraints of OPT1. In this case, due to (\ref{eq:mod_Lit}),
it follows that 
\[
q_{b}r_{p}(0){\cal P}_{D}\leq\lambda_{p}\leq r_{p,max}q_{b}{\cal P}_{D}+q_{b}\left(1-\mathcal{P}_{D}\right)r_{p}\left(0\right),
\]
and, consequently, 
\[
\frac{\lambda_{p}}{{\cal P}_{D}r_{p,max}+\left(1-{\cal P}_{D}\right)r_{p}\left(0\right)}\leq q_{b}\leq\frac{\lambda_{p}}{{\cal P}_{D}r_{p}\left(0\right)}.
\]
 Taking into account that $q_{b}\leq1,$ (\ref{eq:space}) follows.
Conversely, it is assumed that (\ref{eq:space}) holds. By choosing
the vectors 
\[
q^{1}(1,0\left|b\right.)=1,\ q^{1}\left(s,i\left|b\right.\right)=0\ \mbox{otherwise},
\]
and
\[
\sum_{s\in{\cal S}}q^{1}(s,0\left|e\right.)=0,\ \sum_{s\in{\cal S}}\sum_{i\in{\cal I}_{s}}q^{1}\left(s,i\left|e\right.\right)=1,
\]
Eq. (\ref{eq:mod_Lit}) results to $\lambda_{p}^{1}=q_{b}\mathcal{P}_{D}r_{p}(0).$
Similarly, if $(s^{*},i^{*})$ satisfies $r_{p}\left(s^{*},i^{*}\right)=\max_{s,i}\left\{ r_{p}\left(s,i\right)\right\} ,$
by choosing the vectors 
\[
q^{2}(s^{*},i^{*}\left|b\right.)=1,\ q^{2}\left(s,i\left|b\right.\right)=0\ \mbox{otherwise}
\]
and 
\[
\sum_{s\in{\cal S}}q^{2}(s,0\left|e\right.)=1,\ \sum_{s\in{\cal S}}\sum_{i\in{\cal I}_{s}}q^{2}\left(s,i\left|e\right.\right)=0
\]
results to 
\[
\lambda_{p}^{2}=q_{b}\mathcal{P}_{D}r_{p,max}+q_{b}\left(1-\mathcal{P}_{D}\right)r_{p}\left(0\right).
\]
Since by (\ref{eq:mod_Lit}) it holds $\lambda_{p}^{1}\leq\lambda_{p}\leq\lambda_{p}^{2}$
there is an $\alpha$ such that $\alpha\lambda_{p}^{1}+(1-\alpha)\lambda_{p}^{2}=\lambda_{p}$
with $0\leq\alpha\leq1$. Hence, the vectors 
\[
q(s,i\left|b\right.)=\alpha q^{1}\left(s,i\left|b\right.\right)+(1-\alpha)q^{2}(s^{*},i^{*}\left|b\right.)
\]
and
\[
q(s,i\left|e\right.)=\alpha q^{1}\left(s,i\left|e\right.\right)+(1-\alpha)q^{2}(s,i\left|e\right.)
\]
satify the constraints of OPT1.

\bibliographystyle{IEEEtran}
\bibliography{References,IEEEabrv}

\newcommand{\noopsort}[1]{} \newcommand{\printfirst}[3]{#1}
  \newcommand{\singleletter}[1]{#1} \newcommand{\switchargs}[4]{#3#1}
\begin{thebibliography}{10}
\providecommand{\url}[1]{#1}
\csname url@samestyle\endcsname
\providecommand{\newblock}{\relax}
\providecommand{\bibinfo}[2]{#2}
\providecommand{\BIBentrySTDinterwordspacing}{\spaceskip=0pt\relax}
\providecommand{\BIBentryALTinterwordstretchfactor}{4}
\providecommand{\BIBentryALTinterwordspacing}{\spaceskip=\fontdimen2\font plus
\BIBentryALTinterwordstretchfactor\fontdimen3\font minus
  \fontdimen4\font\relax}
\providecommand{\BIBforeignlanguage}[2]{{%
\expandafter\ifx\csname l@#1\endcsname\relax
\typeout{** WARNING: IEEEtran.bst: No hyphenation pattern has been}%
\typeout{** loaded for the language `#1'. Using the pattern for}%
\typeout{** the default language instead.}%
\else
\language=\csname l@#1\endcsname
\fi
#2}}
\providecommand{\BIBdecl}{\relax}
\BIBdecl

\bibitem{FCC}
``Report of the spectrum efficiency working group,'' FCC Spectrum Policy Task
  Force, Tech. Rep. 02-135, 2002.

\bibitem{Mitola}
J.~Mitola, ``Cognitive radio: An integrated agent architecture for software
  defined radio,'' Ph.D. dissertation, KTH, Stockholm, Sweden, 2000.

\bibitem{A:Haykin}
S.~Haykin, ``Cognitive radio: Brain-empowered wireless communications,''
  \emph{IEEE J. Sel. Areas Commun.}, vol.~23, no.~2, pp. 201--220, Feb. 2005.

\bibitem{A:DSA1}
I.~F. Akyildiz, W.-Y. Lee, M.~C. Vuran, and S.~Mohanty, ``Next
  generation/dynamic spectrum access/cognitive radio wireless networks: A
  survey.'' \emph{Comput. Netw.}, vol.~50, no.~13, pp. 2127--2159, Sept. 2006.

\bibitem{DSA2}
Q.~Zhao and B.~Sadler, ``A survey of dynamic spectrum access,'' \emph{IEEE
  Signal Processing Magazine}, vol.~24, no.~3, pp. 79--89, May 2007.

\bibitem{Peng}
C.~Peng, H.~Zheng, and B.~Y. Zhao, ``Utilization and fairness in spectrum
  assignment for opportunistic spectrum access,'' \emph{ACM/Springer MONET},
  vol.~11, no.~4, Aug. 2006.

\bibitem{Chen-2008-ID52}
Y.~Chen, Q.~Zhao, and A.~Swami, ``Joint design and separation principle for
  opportunistic spectrum access in the presence of sensing errors,'' \emph{IEEE
  Trans. Inf. Theory}, vol.~54, pp. 2053--2071, Jan 2008.

\bibitem{Urgaonkar-2009-ID47}
R.~Urgaonkar and M.~J. Neely, ``Opportunistic scheduling with reliability
  guarantees in cognitive radio networks,'' \emph{IEEE Trans. Mobile Comput.},
  vol.~8, pp. 766--777, Jan 2009.

\bibitem{Goldsmith-2009-ID359}
A.~Goldsmith, S.~A. Jafar, I.~Maric, and S.~Srinivasa, ``Breaking spectrum
  gridlock with cognitive radios: An information theoretic perspective,'' in
  \emph{Proc. IEEE}, vol.~97, Jan 2009, pp. 894--914.

\bibitem{Simeone-2008-ID472}
O.~Simeone, I.~Stanojev, S.~Savazzi, Y.~Bar-Ness, U.~Spagnolini, and
  R.~Pickholtz, ``Spectrum leasing to cooperating secondary ad hoc networks,''
  \emph{IEEE J. Sel. Areas Commun.}, vol.~26, pp. 203--213, Jan 2008.

\bibitem{Simeone-2007-ID29}
O.~Simeone, Y.~Bar-Ness, and U.~Spagnolini, ``Stable throughput of cognitive
  radios with and without relaying capability,'' \emph{IEEE Trans. Commun.},
  vol.~55, pp. 2351--2360, Jan 2007.

\bibitem{Krikidis-2009-ID233}
I.~Krikidis, J.~Laneman, J.~Thompson, and S.~Mclaughlin, ``Protocol design and
  throughput analysis for multi-user cognitive cooperative systems,''
  \emph{IEEE Trans. Wireless Commun.}, vol.~8, pp. 4740--4751, Jan 2009.

\bibitem{Kompella2011}
S.~Kompella, G.~D. Nguyen, J.~Wieselthier, and A.~Ephremides, ``Stable
  throughput tradeoffs in cognitive shared channels with cooperative
  relaying,'' in \emph{Proc. IEEE INFOCOM}, 2011, pp. 1961--1969.

\bibitem{Neely_cooper}
R.~Urgaonkar and M.~Neely, ``Opportunistic cooperation in cognitive femtocell
  networks,'' \emph{IEEE J. Sel. Areas Commun.}, vol.~30, no.~3, pp. 607 --616,
  April 2012.

\bibitem{B:Altman}
E.~Altman, \emph{Constrained Markov Decision Processes}.\hskip 1em plus 0.5em
  minus 0.4em\relax Chapman \& Hall/CRC, 1999.

\bibitem{B:Neely_book}
M.~J. Neely, \emph{Stochastic Network Optimization with Application to
  Communication \& Queueing Systems}.\hskip 1em plus 0.5em minus 0.4em\relax
  Morgan \& Claypool, Aug. 2010.

\bibitem{Srikant2013}
R.~Srikant and L.~Ying, \emph{Communication Networks: An Optimization, Control
  and Stochastic Networks Perspective}.\hskip 1em plus 0.5em minus 0.4em\relax
  Cambridge University Press, 2013.

\bibitem{A:Mo}
J.~Mo and J.~Walrand, ``Fair end-to-end window-based congestion control,''
  \emph{IEEE/ACM Trans. Netw.}, vol.~8, no.~5, pp. 556--567, 2000.

\bibitem{Lo}
B.~F. Lo, ``A survey of common control channel design in cognitive radio
  networks,'' \emph{Ph. Commun.}, vol.~4, no.~1, pp. 26--39, 2011.

\bibitem{Ephrem}
J.~E. Wieselthier, G.~D. Nguyen, and A.~Ephremides, ``Energy-efficient
  broadcast and multicast trees in wireless networks,'' \emph{Mob. Networks and
  Appl.}, vol.~7, pp. 481--492, 2002.

\bibitem{Papadimitriou}
I.~Papadimitriou and L.~Georgiadis, ``Minimum energy broadcasting in multihop
  wireless networks using a single broadcast tree,'' \emph{Mob. Networks and
  Appl.}, vol.~11, pp. 361--375, 2006.

\bibitem{Boyd}
S.~Boyd, N.~Parikh, E.~Chu, B.~Peleato, and J.~Eckstein, ``Distributed
  optimization and statistical learning via the alternating direction method of
  multipliers,'' \emph{Foundations and Trends in Machine Learning}, vol.~3,
  no.~1, pp. 1--122, 2011.

\bibitem{D.P.Bertsekas_book}
D.~Bertsekas, \emph{Constrained Optimization and Lagrange Multiplier
  Methods}.\hskip 1em plus 0.5em minus 0.4em\relax 2nd ed. Belmont, MA: Athena
  Scientific, 1996.

\bibitem{BertsekasTsitsiklis_book}
D.~Bertsekas and N.~J. Tsitsiklis, \emph{Parallel and Distributed Computation:
  Numerical Methods}.\hskip 1em plus 0.5em minus 0.4em\relax 2nd ed. Belmont,
  MA: Athena Scientific, 1999.

\end{thebibliography}

\end{document}